\documentclass[lettersize,journal]{IEEEtran}

\usepackage{cite}
\usepackage{amsmath,amssymb,amsfonts}
\usepackage{algorithmic}
\usepackage{graphicx}
\usepackage{siunitx}
\usepackage{textcomp}
\usepackage{xcolor}
\usepackage{subcaption}
\usepackage{booktabs}
\usepackage{multirow}
\usepackage{comment}
\usepackage{booktabs}
\usepackage{siunitx}
\usepackage{xspace}

\usepackage{xurl} 
\usepackage[hidelinks]{hyperref} 
\newcommand{\mumap}{$\mu$-map\xspace}
\newcommand{\wdtnv}{w-{\rm dTNV}\xspace}
\newcommand{\wtnv}{w-{\rm TNV}\xspace}
\newcommand{\dtv}{{\rm dTV}\xspace}

\def\BibTeX{{\rm B\kern-.05em{\sc i\kern-.025em b}\kern-.08em
    T\kern-.1667em\lower.7ex\hbox{E}\kern-.125emX}}

\begin{document}
\title{Weighted Directional Total Nuclear Variation for Joint $^{90}$Y PET/SPECT Reconstruction with CTAC-derived Guidance}

\author{S Porter, \IEEEmembership{Student Member, IEEE}, D Deidda, \IEEEmembership{Senior Member, IEEE}, D R McGowan, J Anton-Rodriguez, S Arridge and K Thielemans, \IEEEmembership{Senior Member, IEEE} 
\thanks{Submitted for review on 19 April 2026.} 
\thanks{This work was supported in part by National Physical Laboratory through the National Measurement System of the Department for Science, Innovation and Technology (NPL grant PO496581) and by the Engineering and Physical Sciences Research Council (EPSRC) through an Industrial CASE studentship (EPSRC grant EP/W522077/1) and Mediso Medical Imaging Systems. Software used in this project is maintained by CCP SyneRBI (UKRI support: EPSRC grant EP/T026693/1), CCPi (EPSRC grant EP/T026677/1) and the Science and Technology Facilities Council through the Digital Research Infrastructure programme.}
\thanks{S. Porter is with the Institute of Nuclear Medicine and Hawkes Institute, University College London (UCL), and the Medical, Marine and Nuclear Division, The National Physical Laboratory (NPL) (e-mail: sam.porter.18@ucl.ac.uk).}
\thanks{D. Deidda is with Medical, Marine and Nuclear Division, NPL.}
\thanks{D.R. McGowan is with Oxford University Hospitals NHS Foundation Trust and University of Oxford.}
\thanks{J. Anton-Rodriguez is with The Christie NHS Trust.}
\thanks{S. Arridge is with the Centre for Inverse Problems and Hawkes Institute, UCL.}
\thanks{K. Thielemans is with the Institute of Nuclear Medicine and Hawkes Institute, UCL.}
\thanks{Special thanks go to Xavier Loizeau (NPL) for assistance with the statistical testing and Efstathios Varzakis for the PhantomGen software used to construct NEMA VOIs.}}

\markboth{IEEE Transactions on Radiation and Plasma Medical Sciences,~Vol.~XX, No.~X, February~2026}%
{Weighted Directional Total Nuclear Variation for Joint $^{90}$Y PET/SPECT Reconstruction with CTAC-derived Guidance}

\maketitle

\begin{abstract}
Quantitative post-treatment activity imaging is required for personalised dosimetry after $^{90}$Y selective internal radiation therapy (SIRT). $^{90}$Y PET offers high spatial resolution but has extremely low-count noise, whereas $^{90}$Y bremsstrahlung SPECT has higher count statistics but suffers from severe blur, scatter, and septal penetration. Since both modalities image the same microsphere distribution, there is potential to exploit the strong physical coupling by using joint (synergistic) reconstruction. In addition, as PET and SPECT data are acquired together with CT for attenuation correction (CTAC), anatomical guidance is available as well.   We propose weighted directional total nuclear variation (\wdtnv), a joint variational regulariser for coupled PET/SPECT reconstruction with CTAC-guided anisotropy. \wdtnv penalises the nuclear norm of a dual-modality Jacobian to promote co-located, geometry-consistent edges without enforcing intensity correlation. CTAC guidance is incorporated via a directional operator acting on PET and SPECT gradients, computed from the CTAC-derived attenuation map $\mu$, enabling efficient per-voxel spectral computations. To mitigate scale disparity between $^{90}$Y PET and SPECT, we use data-driven modality normalisation computed from preliminary reconstructions. We compare \wdtnv on a NEMA IEC phantom using 20 bootstrapped PET noise realisations and on 9 post-SIRT patients (45 lesions) against comparators: an isotropic dual-modality baseline, a directional non-synergistic baseline, and a sequential hybrid kernel method (SHKEM). In the phantom, \wdtnv improves recovery coefficients relative to \dtv and \wtnv, and improves recovery relative to SHKEM for the smallest spheres. In patients, \wdtnv resulted in higher tumour-to-background ratios than SHKEM at comparable background. Overall, these results indicate that CTAC-guided synergistic variational coupling can improve phantom lesion recovery and clinical lesion contrast, with potential implications for quantitative low-count $^{90}$Y imaging, offering a practical route towards more reliable post-SIRT activity imaging for personalised dosimetry through more stable post-treatment activity estimates.
\end{abstract}

\begin{IEEEkeywords}
Yttrium-90, PET-SPECT, joint reconstruction, radioembolisation, multi-modal imaging, anatomical guidance, image regularisation, quantification
\end{IEEEkeywords}

\section{Introduction}
\label{sec:introduction}

Post-treatment verification of $^{90}$Y selective internal radiation therapy (SIRT) requires quantitative activity imaging for personalised dosimetry \cite{levillainInternationalRecommendationsPersonalised2021}. Pre-treatment planning based on $^{99\mathrm{m}}$Tc-MAA frequently poorly predicts the microsphere distribution, motivating post-treatment imaging for treatment verification \cite{hasteCorrelationTechnetium99mMacroaggregated2017}.

Two complementary $^{90}$Y modalities are clinically available. PET detects (rare) internal pair-production, with higher spatial resolution but extremely low count statistics \cite{selwynNewInternalPair2007}. Bremsstrahlung SPECT provides substantially higher overall counts after correction ($\sim 15\times$ for our data) but is degraded by severe blur, septal penetration, and scatter due to the continuous emission spectrum \cite{rongDevelopmentEvaluationImproved2012}. In clinical practice, both modalities are typically acquired alongside CT used for attenuation correction and anatomical localisation (CTAC). However, CTAC provides structural guidance rather than direct activity information, and PET and SPECT are still generally reconstructed independently.

Deidda \textit{et al.} demonstrated improved lesion contrast using a sequential hybrid kernel expectation maximisation (SHKEM) approach \cite{deiddaTripleModalityImage2023, deiddaHybridPETMRListmode2019}, but SHKEM remains sensitive to iteration and noise amplification. This motivates a truly joint formulation with a coupled prior $\mathcal{R}(\mathbf{z})$ that promotes shared structure while allowing modality-specific contrast \cite{knollJointMRPETReconstruction2017, arridgeOverviewSynergisticReconstruction2021}. 

\subsection{Contributions}
This work makes three contributions:
\begin{enumerate}
    \item We introduce weighted directional total nuclear variation (\wdtnv), a joint PET/SPECT regulariser based on total nuclear variation \cite{knollJointMRPETReconstruction2017} with CTAC guidance incorporated as a directional operator acting on PET and SPECT gradients using a CTAC-derived attenuation map ($\mu$-map).
    \item We evaluate the repeatability and quantitative performance of \wdtnv against the sequential hybrid kernel method of Deidda \textit{et al.} \cite{deiddaTripleModalityImage2023} on a NEMA IEC phantom with bootstrapped PET noise realisations and on post-SIRT patient data.
    \item We perform an ablation study where \wdtnv is compared against isotropic weighted total nuclear variation (\wtnv) and the non-synergistic directional total variation (\dtv).
\end{enumerate}

\section{Background}
\subsection{Physics of $^{90}$Y Imaging}
\label{subsec:physics}
$^{90}$Y decays predominantly by $\beta^{-}$ emission to $^{90}$Zr (half-life 64.1\,h). Coulomb interactions of the emitted electrons generate a continuous bremsstrahlung spectrum enabling SPECT imaging, but the continuum and high-energy component induce substantial penetration, scatter, and system blur \cite{minarikEvaluationQuantitative90Y2008}. A rare internal pair-production branch results in positron annihilation photons enabling PET imaging, providing higher spatial resolution but very low count statistics \cite{selwynNewInternalPair2007}.

\subsection{Related Work}\label{subsec:related}
PET and SPECT reconstruction can be formulated as inverse problems, but are generally ill-posed, especially for low-count data. Iterative expectation--maximisation methods such as MLEM and OSEM \cite{dempsterMaximumLikelihoodIncomplete1977, sheppMaximumLikelihoodReconstruction1982, hudsonAcceleratedImageReconstruction1994} converge to noise-dominated solutions without explicit regularisation. Early stopping acts as an implicit regulariser \cite{veklerovStoppingRuleMLE1987} but introduces bias and degrades quantitative accuracy \cite{jaskowiakInfluenceReconstructionIterations2005}.

Bayesian formulations make the regularisation explicit via a prior term $\mathcal{R}(x)$ in the maximum a posteriori (MAP) objective. Edge-preserving priors such as total variation (TV) \cite{rudinNonlinearTotalVariation1992} suppress noise whilst maintaining sharp transitions \cite{jonssonTotalVariationRegularizationPositron1998}. Anatomically guided priors exploit structural information from CT or MR to further preserve legitimate boundaries \cite{ehrhardtMultimodalityImagingStructurepromoting2020a}. Examples include the Bowsher prior \cite{bowsherUtilizingMRIInformation2004}, which encourages similarity between anatomically neighbouring voxels, and directional TV \cite{ehrhardtPETReconstructionAnatomical2016}, which aligns gradient orientations between the emission image and an anatomical reference. Kernel methods \cite{wangPETImageReconstruction2015, hutchcroftAnatomicallyaidedPETReconstruction2016} achieve similar effects by reparameterising the image in a feature space derived from anatomical or functional structure, embedding guidance directly into the forward model. Hybrid kernel methods \cite{deiddaHybridKernelisedExpectation2022} incorporate features from both a guidance image and the current iterate. Sequential reconstruction pipelines where one modality is reconstructed to be used as guidance for another have also been introduced \cite{deiddaTripleModalityImage2023, marquisTheranosticSPECTReconstruction2021}.

Truly synergistic methods reconstruct multiple modalities simultaneously and can enforce structural coupling through a joint regulariser $\mathcal{R}(\mathbf{z})$. This approach has been extensively investigated for PET-MR \cite{ehrhardtJointReconstructionPETMRI2015, knollJointMRPETReconstruction2017, mehranianNonconvexJointsparsityRegularization2016}.
For a comprehensive review of synergistic reconstruction, see Arridge \textit{et al.} \cite{arridgeOverviewSynergisticReconstruction2021}.

To our knowledge, truly joint reconstruction has not previously been evaluated for the specific setting of post-treatment $^{90}$Y PET/SPECT, where both functional modalities image the same microsphere distribution but differ substantially in noise, resolution, and artefact characteristics, while CTAC provides anatomical rather than activity information. This work develops a joint variational framework overcoming these limitations.

\section{Methods}
\label{sec:methods}

\subsection{Synergistic Methods}\label{subsec:synergistic}

This section presents our synergistic reconstruction framework based on directional total nuclear variation (TNV). For comparison, we also describe the sequential hybrid kernel method (SHKEM) of Deidda et al., which represents the current state-of-the-art for $^{90}$Y PET/SPECT reconstruction.

\subsubsection{MAP Reconstruction for Synergy}
\label{sec:recon_basics}

In emission tomography, the unknown tracer distribution is discretised on a voxel grid and estimated from Poisson-distributed projection data. Let $x \in \mathbb{R}_+^{J}$ denote the activity image and $y \in \mathbb{N}^{I}$ the measured counts. Under the standard affine forward model
\begin{equation}
  \bar y = Ax + b,
\end{equation}
the data are modelled as independent Poisson random variables, $y_i \sim \mathrm{Poisson}(\bar y_i)$. Reconstruction is commonly posed as a maximum a posteriori (MAP) problem,
\begin{equation}
  \hat x
  =\arg\min_{x \ge 0}\left\{\mathcal{O}(x)\right\} =\arg\min_{x \ge 0}
  \Bigl\{
    \mathcal{D}(y \mid \bar y(x)) + \mathcal{R}(x)
  \Bigr\},
\end{equation}
where $\mathcal{D}$ is the Poisson negative log-likelihood,
\begin{equation}
\mathcal{D}(y\mid \bar y)=\sum_{i=1}^I \left(\bar y_i - y_i\log \bar y_i\right) + \mathrm{const.},
\end{equation}
and $\mathcal{R}$ is a regulariser encoding prior information.

For $M$ modalities, let $z_m \in \mathbb{R}_+^N$ denote the image for modality $m$ represented on a shared reference grid comprising $N$ voxels, with corresponding data $y_m \in \mathbb{N}^{I_m}$. Here $N$ need not coincide with the native image size $J_m$ of modality $m$, so that the mapping to the shared grid may include both spatial warping and a change of resolution. Once the registration transform is fixed, let 
$W_m : \mathbb{R}^{J_m} \to \mathbb{R}^N$ denote the resulting discrete linear resampling operator from the native grid of modality $m$ to the shared grid, and let 
$W_m^\ast : \mathbb{R}^N \to \mathbb{R}^{J_m}$ denote its discrete adjoint with respect to the chosen image-space inner products. The adjoint maps shared-grid variables back to the native image grid in the data-fidelity term prior to application of the system matrix $A_m$. The predicted data are therefore
\begin{equation}
  \bar y_m = A_m W_m^\ast z_m + b_m
\end{equation}
and the joint MAP estimator takes the form
\begin{equation}
  \hat{\mathbf z}
  =
  \arg\min_{\mathbf z \ge 0}
  \left\{
    \sum_{m=1}^M
    \mathcal{D}_m\bigl(y_m \mid A_m W_m^\ast z_m + b_m\bigr)
    +
    \mathcal{R}(\mathbf z)
  \right\},
\end{equation}
where $\mathbf z = (z_1,\dots,z_M)$ and $\mathcal{R}$ couples the modality images through their shared-grid representations.

\subsubsection{Total Nuclear Variation}\label{subsubsec:tnv}

We build on Total Nuclear Variation (TNV) \cite{rigieJointReconstructionMultichannel2015, holtTotalNuclearVariation2014}, a vector-valued extension of TV that promotes collocated and aligned edges across modalities by penalising the nuclear norm of the joint Jacobian. Let $\nabla$ denote a discrete forward-difference operator that returns, at each voxel $n$, a vector of directional finite differences in $d$ prescribed directions (defined below). Hence, for any image $u$, $(\nabla u)_n \in \mathbb{R}^{d}$ and $\|(\nabla u)_n\|_2$ denotes the Euclidean norm in this $d$-dimensional directional-difference space.

Define the multi-modality Jacobian for TNV as
\begin{equation}
\mathcal{G}^{\mathrm{TNV}}_n(\mathbf{z})=
\bigl[(\nabla z_1)_n \;|\; \cdots \;|\; (\nabla z_M)_n\bigr]\in\mathbb{R}^{d\times M},
\end{equation}
and the TNV prior
\begin{equation}
\mathcal{R}_{\mathrm{TNV}}(\mathbf{z})
=
\sum_{n=1}^N \|\mathcal{G}^{\mathrm{TNV}}_n(\mathbf{z})\|_*,
\qquad
\|Y\|_*=\sum_{\ell}\sigma_\ell(Y),
\end{equation}
where $\sigma_\ell(Y)$ are the singular values of matrix $Y$ and $\|Y\|_*$ is the nuclear norm. Since $\|\cdot\|_*$ is non-smooth, we use a smooth surrogate that is differentiable everywhere \cite{lewisDerivativesSpectralFunctions1996},
\begin{equation}
\Phi_\epsilon(Y)=\sum_{\ell=1}^{\min(d,M)} g\bigl(\sigma_\ell(Y)\bigr),
\qquad
g(s)=\sqrt{s^2+\epsilon^2},
\end{equation}
so that
\begin{equation}
g'(s)=\frac{s}{\sqrt{s^2+\epsilon^2}},
\qquad
\frac{\partial}{\partial \sigma_\ell}\Phi_\epsilon(Y)=g'\!\bigl(\sigma_\ell(Y)\bigr).
\end{equation}
This enables efficient gradient-based updates via closed-form derivatives. Let $Y = U\,\mathrm{diag}(\sigma)\,V^\top$ be the singular value decomposition (SVD) of $Y$, where $U\in\mathbb{R}^{d\times r}$, $V\in\mathbb{R}^{M\times r}$, $r=\mathrm{rank}(Y)$, and $\sigma\in\mathbb{R}_+^{r}$ collects the nonzero singular values. For the spectral function $F(Y)=\sum_{\ell} g(\sigma_\ell(Y))$, with differentiable $g$, the gradient with respect to $Y$ is
\begin{equation}
\nabla_Y F(Y)=U\,\mathrm{diag}\bigl(g'(\sigma)\bigr)\,V^\top,
\label{eq:spectral_grad}
\end{equation}
which is a standard expression for differentiable spectral functions; see, e.g., \cite{lewisDerivativesSpectralFunctions1996}.

\subsubsection{Weighted directional TNV with CT-based guidance}\label{subsubsec:wdtnv}

We now adapt TNV to the proposed \emph{weighted directional TNV} (\wdtnv), which incorporates anatomical guidance from CTAC while directly coupling only the emission modalities. Two additional ingredients are introduced. First, positive modality weights $w_m>0$ rescale the Jacobian columns to account for differing native intensity scales. Second, directional operators $D_{m,n}$ anisotropically modulate gradient components using a CTAC-derived guidance field.

Anatomical information is incorporated through the CTAC-derived attenuation map, denoted by $\mu$ (the ``$\mu$-map''). Specifically, $\mu$ is reconstructed from the low-dose CTAC and resampled to the reconstruction grid.

It is possible to include $\mu$ as a third coupled channel in the TNV Jacobian. However, this presents difficulties due to the very different local scale behaviour between the emission and attenuation modalities. Although one could introduce an additional scaling factor for the $\mu$-channel, tuning PET and SPECT alone is already non-trivial. More importantly, a single global weighting of $\mu$ does not yield satisfactory behaviour voxelwise: in regions where emission gradients are weak, even a modest $\mu$ contribution can become the dominant column of the local Jacobian, whereas in regions with strong emission edges the same scaling may render $\mu$ effectively irrelevant. Instead, we propose to use $\mu$ only to steer the functional regulariser anisotropically via a directional operator acting on the PET and SPECT gradients, previously used in DTV \cite{ehrhardtPETReconstructionAnatomical2016}, with explicit control over anatomical influence through the stabilised direction field (via $\eta$ and $\{\gamma_m\}$, defined below). Retaining a two-channel Jacobian also yields a $2\times 2$ Gram matrix, enabling efficient closed-form evaluation of singular values.

We therefore propose \emph{weighted directional TNV} (\wdtnv). We define the stabilised $\mu$-gradient direction
\begin{equation}
\xi_n=\frac{(\nabla \mu)_n}{\|(\nabla \mu)_n\|_2+\eta},
\qquad
\eta = 0.05 \cdot \max_n \|(\nabla \mu)_n\|_2.
\end{equation}
We pre-smooth $\mu$ with only a very small blurring kernel ($\mathrm{FWHM}=$ \SI{2}{\milli\metre}) to retain a high guidance edge-resolution; robustness is instead obtained by the stabiliser $\eta$, which aims to prevent amplification of small-magnitude $\mu$ differences in homogeneous regions. Note that $\xi_n\in\mathbb{R}^d$ lives in the same directional-difference space as $(\nabla z_m)_n$. For each modality $m$, define a pointwise directional operator acting on a gradient field $\nabla z_m\in\mathbb{R}^{N\times d}$ as
\begin{equation}
D_{m,n}(\nabla z_m)_n = (I_d-\gamma_m\,\xi_n\xi_n^\top)(\nabla z_m)_n,
\qquad
\gamma_m\in[0,1].
\end{equation}
With $\gamma_m=1$ (used in this work for both modalities), $D_{m,n}$ applies an anisotropic shrinkage along $\xi_n$, down-weighting gradient components parallel to the local $\mu$-edge normal direction. Since $\xi_n$ is stabilised and generally has norm less than one, this operation should be interpreted as anisotropic shrinkage rather than complete projection along the $\mu$-gradient direction. Consequently, gradient components aligned with the $\mu$ edge contribute less to the penalty, making $\mu$-consistent edges preferable.

At voxel $n$ we form the dual-modality Jacobian
\begin{equation}
\mathcal{G}^{\wdtnv}_n(\mathbf{z})=
\left[\, w_{1}D_{1,n}(\nabla z_1)_n \;\Big|\; w_{2}D_{2,n}(\nabla z_2)_n \right]\in\mathbb{R}^{d\times 2}.
\end{equation}
The weights $w_m$ balance native PET/SPECT scales in $^{90}$Y imaging \cite{pasciakRadioembolizationDynamicRole2014}; otherwise the larger-scale modality dominates $\sigma_\ell(\mathcal{G}^{\wdtnv}_n)$. We use spatially uniform, data-driven weights
\begin{equation}
w_m
=
\frac{\omega_m}{q_{0.99}\!\bigl(z_m^{\mathrm{init}}\!\mid_{\Omega'}\bigr)-q_{0.01}\!\bigl(z_m^{\mathrm{init}}\!\mid_{\Omega'}\bigr)},
\end{equation}
where $q_p(\cdot\mid_{\Omega'})$ denotes the $p$th percentile computed over a foreground mask (CTAC-based) $\Omega'$, $x_m^{\mathrm{init}}$ is an unregularised early-stopped OSEM reconstruction on the native grid of modality $m$, $z_m^{\mathrm{init}}:=W_m x_m^{\mathrm{init}}$ is its representation on the shared grid, and $\omega_m$ are user-specified global regularisation weights. Percentile scaling was used as a robust dynamic-range estimate, reducing sensitivity to outliers. The constants used for directional stabilisation, smoothing, and weighting were selected empirically during method development and then held fixed; they were not tuned separately for individual evaluation cases.

\subsubsection{Choice of Directions}
Let $\Delta = (\Delta_x,\Delta_y,\Delta_z)$ denote voxel sizes (SI units). We define a set of forward offsets from a nine-direction half-stencil comprising three axial face directions and six edge directions (excluding corners),
\begin{equation}
\footnotesize
\mathcal{E}_1=\{(1,0,0),(0,1,0),(0,0,1)\}\cup\{(\pm1,1,0),(\pm1,0,1),(0,1,\pm1)\},
\end{equation}
and use both one- and two-voxel steps, $\mathcal{E}=\mathcal{E}_1\cup\{2e:e\in\mathcal{E}_1\}$, yielding $d=18$ directions. For an image $u$, the discrete directional gradient $(\nabla u)_n\in\mathbb{R}^{d}$ is the stack of scaled forward differences
\begin{equation}
(\nabla u)_n :=
\left(
\frac{u_{n+e}-u_n}{\|e\|_{\Delta}}
\right)_{e\in\mathcal{E}},
\end{equation}
where $\mathcal{E}$ is the set of integer voxel offsets described above, $e=(e_x,e_y,e_z)$, and
\begin{equation}
\|e\|_{\Delta} := \sqrt{(e_x\Delta_x)^2+(e_y\Delta_y)^2+(e_z\Delta_z)^2}
\end{equation}
is the corresponding physical step length, reducing to the usual Euclidean offset length for isotropic unit voxels. Differences crossing the image boundary were evaluated using Neumann boundary conditions. Extended Jacobians have been demonstrated to improve reconstruction using TNV \cite{holtTotalNuclearVariation2014}. In addition, the increased step length was chosen to more appropriately handle lower SPECT resolution and possible registration errors.

\subsubsection{Final Form of the Synergistic Penalty}
Summing the local penalties over all voxels, with weighting by the voxel volume, gives
\begin{equation}
\label{eq:wdtnv_jacobian}
\mathcal{R}_{\wdtnv}(\mathbf{z})=V_{\mathrm{vox}}\sum_{n=1}^N \Phi_\epsilon\!\bigl(\mathcal{G}^{\wdtnv}_n(\mathbf{z})\bigr).
\end{equation}
We set $\epsilon$ relative to initial joint-gradient magnitudes. With $\Omega'$ as defined above,
\begin{equation}
\epsilon:=0.05 \cdot \,\mathrm{median}_{n\in\Omega'}\,\sigma_1(\mathcal{G}^{\wdtnv}_n(\mathbf{z}^{\mathrm{init}})),
\end{equation}
where $\mathbf{z}^{\mathrm{init}} := (z_1^{\mathrm{init}},\dots,z_M^{\mathrm{init}})$. This was chosen empirically to smooth low-count noise. In the unsmoothed limit, the coupling when one channel provides no structural information reduces pointwise to a single-modality TV-type penalty on the non-uniform modality: if $D_{2,n}(\nabla z_2)_n=0$ (e.g.\ $z_2$ is locally uniform), then
\begin{equation}
\|\mathcal{G}^{\wdtnv}_n(\mathbf{z})\|_* = w_1\|D_{1,n}(\nabla z_1)_n\|_2,
\end{equation}
i.e.\ \wdtnv\ (and likewise \wtnv\ with $D_{m,n}=I_d$) reduces locally to a single-modality TV-type penalty in this limiting sense.

\subsubsection{Ablation comparators}
\label{subsubsec:ablations}

To isolate the contributions of synergistic PET/SPECT coupling and CTAC-derived directional guidance, we compared \wdtnv with two variational ablations. First, isotropic weighted TNV (\wtnv) removes the directional CTAC operator while retaining dual-modality coupling. This corresponds to setting
\begin{equation}
D_{m,n}=I_d,\qquad m\in\{\mathrm{PET},\mathrm{SPECT}\},
\end{equation}
in Eq.~\eqref{eq:wdtnv_jacobian}, giving
\begin{equation}
\mathcal{G}^{\wtnv}_n(\mathbf z)
=
\left[
w_1(\nabla z_1)_n
\;\middle|\;
w_2(\nabla z_2)_n
\right].
\end{equation}
Thus, \wtnv tests the effect of PET/SPECT synergistic coupling without CTAC-guided anisotropy.

Second, directional total variation (\dtv) removes the synergistic PET/SPECT coupling while retaining CTAC-derived directional guidance for PET. It is obtained by applying the same directional operator to the PET gradient only:
\begin{equation}
\mathcal{R}_{\dtv}(z_{\mathrm{PET}})
=
V_{\mathrm{vox}}
\sum_{n=1}^N
\Phi_\epsilon
\left(
w_{\mathrm{PET}}
D_{\mathrm{PET},n}
(\nabla z_{\mathrm{PET}})_n
\right),
\end{equation}
which, in the single-channel case, reduces to a smoothed directional TV penalty. Thus, \dtv tests the effect of CTAC-guided anisotropy without PET/SPECT coupling.

All ablation reconstructions used the same forward models, registration, optimisation scheme, foreground mask, dynamic-range normalisation, and CTAC-derived direction field as \wdtnv, with only the regulariser structure changed.

\subsubsection{Optimisation}
\label{subsubsec:optimisation}

We minimise the joint objective using a preconditioned stochastic variance-reduced gradient (SVRG) algorithm \cite{johnsonAcceleratingStochasticGradient2013}. We implement subset SVRG by decomposing the objective as
\begin{equation}
\mathcal{O}(\mathbf{z})
=
\sum_{s=1}^S \left [
\mathcal{D}_s(\mathbf{z})
\;+\;
\frac{1}{S}\mathcal{R}(\mathbf{z})
\right],
\end{equation}
i.e.\ the regulariser is replicated across subsets with weight $1/S$ so that $\sum_{s}(1/S)\mathcal{R}=\mathcal{R}$ and each subset gradient contains a consistent regularisation contribution. Measurements are partitioned into $S$ disjoint subsets. At each iteration $k$, a subset index $s_k$ is sampled uniformly from $\{1,\dots,S\}$ and we form the SVRG gradient estimator
\begin{align}
g^{(k)}
&:=
\nabla \mathcal{O}(\tilde{\mathbf{z}})
+
S\Bigl[\nabla \mathcal{O}_{s_k}(\mathbf{z}^{(k)}) - \nabla \mathcal{O}_{s_k}(\tilde{\mathbf{z}})\Bigr],
\\
\mathcal{O}_s(\mathbf{z})
&:=
\mathcal{D}_s(\mathbf{z}) + \frac{1}{S}\mathcal{R}(\mathbf{z}),
\end{align}
where $\tilde{\mathbf{z}}$ is an anchor point updated once per epoch by evaluating $\nabla\mathcal{O}(\tilde{\mathbf{z}})$ using all subsets. We chose to split subsets by modality, i.e. each $\mathcal{D}_s(\mathbf{z})$ is built from either PET or SPECT data only and within each modality. These choices were motivated by our recent analysis of subset selection in synergistic emission tomography, which found such partitioning to provide a better-behaved optimisation scheme \cite{porter_optimising_2024}. Subsets were formed using staggered (view-interleaved) partitioning. One stochastic epoch corresponds to a full pass over all subsets. Because the full-gradient snapshot is also recomputed once per epoch, each epoch requires approximately two full data passes. This estimator is then used within a projected, diagonally preconditioned gradient update
\begin{equation}
\mathbf{z}^{(k+1)}
=
\left\{
\mathbf{z}^{(k)} - \varsigma^{(k)} P^{(k)} g^{(k)}
\right\}_{\ge 0},
\end{equation}
where $\{\cdot\}_{\ge 0}$ denotes componentwise projection onto the nonnegative orthant. 
We use EM-type diagonal scaling \cite{depierroFastEMlikeMethods2001, ahnGloballyConvergentOrdered2002}. For each modality $m$, the data preconditioner is formed on the shared grid as
\begin{equation}
P_{\mathrm{data},m} = z_m \oslash \bigl(W_m A_m^\top \mathbf{1}\bigr),
\end{equation}
where $\oslash$ denotes elementwise division and $\mathbf{1}$ is an all-ones vector in projection space. Thus, $W_m A_m^\top\mathbf{1}$ is the native-grid sensitivity image mapped back onto the shared reconstruction grid.

To incorporate regulariser scaling, we additionally compute a diagonal curvature proxy $P_{\mathcal R}$, a voxel--modality-wise diagonal approximation to the regulariser Hessian based on the local SVD singular-value weights of $\mathcal G_n(\mathbf z^{(k)})$. For the two-modality case considered here, the required singular values are obtained from the eigenvalues of the local $2\times2$ Gram matrix, avoiding an iterative voxelwise SVD. Since the reconstruction variable $\mathbf z$ is represented on the shared grid, both diagonal scalings are represented on the shared grid before being combined to form $P^{(k)}$. Due to its cost, $P_{\mathcal R}$ is updated once per epoch and held fixed within each epoch. 

Inspired by Erhardt et al. \cite{ehrhardtFastPETReconstruction2025}, we combine the two diagonal
scalings using a floor-regularised Lehmer mean. Let
\begin{equation}
\widetilde P_{\mathrm{data}}
=
\max(P_{\mathrm{data}},p_{\min}),
\qquad
\widetilde P_R
=
\max(P_R,p_{\min}),
\end{equation}
where \(p_{\min}=10^{-6}\) is a small numerical floor. We then define
\begin{equation}
P =
\frac{\widetilde P_{\mathrm{data}}^{q}
      +
      \widetilde P_R^{q}}
     {\widetilde P_{\mathrm{data}}^{q-1}
      +
      \widetilde P_R^{q-1}},
\qquad q=0.1 .
\end{equation}
For \(q=0\), this reduces to harmonic averaging. The choice
\(q=0.1\) gives a slightly less severe, but still conservative,
combination of the data and regularisation scalings. The
positive floor prevents numerical zeros in either scaling from
collapsing the combined preconditioner to zero and freezing
the corresponding update.

The step size was chosen to decay as $\varsigma^{(k)} = \varsigma^{(0)}/(1 + \delta k)$. For all reconstructions, we used $\varsigma^{(0)}=1$ and $\delta=0.005$. Reconstructions were run for 250 epochs (500 full data passes), after which all reconstructions had visually converged. No formal convergence criterion was used.

\subsection{Sequential Hybrid Kernel Method}\label{subsubsec:shkem}

As a comparator, we implemented the sequential hybrid kernel expectation maximisation (SHKEM) method of Deidda \textit{et al.}~\cite{deiddaTripleModalityImage2023}, using the reconstruction order and hyperparameter settings reported as optimal in that work. The sequential order was CT-guided SPECT followed by CT/SPECT-guided PET. Thus, SPECT was reconstructed first using CT-derived anatomical features and hybrid kernel updates. The resulting SPECT reconstruction was then used, together with CT-derived features and the current PET iterate, to construct the PET hybrid kernel.

For modality $m$, the image is represented as
\begin{equation}
    \mathbf{x}_m^{(k)} = \mathbf{K}^{(k)}_{m-1}\boldsymbol{\alpha}_m ,
\end{equation}
where $\mathbf{K}^{(k)}_{m-1}$ is the hybrid kernel constructed from the previous guidance image(s) and the current iterate, and $\boldsymbol{\alpha}_m$ are estimated by minimising the Poisson negative log-likelihood with the composite system matrix $\mathbf{A}_m\mathbf{K}^{(k)}_{m-1}$.

To mitigate noise amplification, we employ frozen kernels~\cite{deiddaHybridKernelisedExpectation2022}, halting kernel updates after 27 PET subiterations and 72 SPECT subiterations. All kernel parameters ($\sigma_v$, $\sigma_d$, $\sigma_\alpha$, neighbourhood size) followed the original implementation from Deidda\,\textit{et al.}

\subsection{Data}
\label{subsec:data}

Data were selected to enable direct comparison with the synergistic reconstruction framework proposed by Deidda\,\textit{et al.} \cite{deiddaTripleModalityImage2023}.

\subsubsection{Phantom Data}
A NEMA IEC Body Phantom with a sphere-to-background ratio of $8$ was scanned at The Christie Hospital (Manchester, UK) using a Siemens Biograph mCT PET/CT system and a GE Discovery~670 SPECT/CT system. PET data were acquired over a total duration of \SI{15}{\hour}, while SPECT data were acquired for \SI{15}{\minute}. From the \SI{15}{\hour} PET acquisition, twenty statistically independent PET datasets equivalent to \SI{15}{\minute} acquisitions were generated using sinogram bootstrapping \cite{markiewiczAssessmentBootstrapResampling2014}. The SPECT energy window was $39.84$--$128.16~\mathrm{keV}$.

\subsubsection{Clinical Data}
Clinical data were acquired at Oxford University Hospitals (OUH) from 10 patients undergoing $^{90}$Y SIRT (1 patient for parameter choice and 9 for testing). PET data were acquired on a GE Discovery~710 PET/CT system, and SPECT imaging was performed on a GE Discovery NM/CT~670 system \cite{rowleyOptimizationImageReconstruction2016}. Informed consent is not necessary for retrospective image reviews of this nature at OUH.

PET data were acquired across two bed positions with \SI{15}{\minute} per bed position and SPECT data were acquired over \SI{15}{\minute} using standard oncology protocols. Energy windows were $50$--$150~\mathrm{keV}$.

\subsection{Forward Models, Corrections, and Discretisation}
\label{subsec:forward_models}

We use an affine forward model $\bar y = Ax + b$ for both modalities, where $A$ denotes the system matrix and $b$ denotes the expected additive correction contribution; the overbar is reserved for the total expected data $\bar y$.

\subsubsection{PET}
PET list-mode data were processed using GE's Duetto PET Toolbox and Siemens' E7 tools to generate normalisation, attenuation, and additive corrections (randoms and scatter), which were treated as fixed during reconstruction~\cite{stearnsRandomCoincidenceEstimation2003, iatrou3DImplementationScatter2006}. Reconstructions use a voxel size of $3.27\times 2.13\times 2.13~\mathrm{mm}^3$ (z,y,x) on a grid of $83\times 255\times 255$. Patient data were acquired with two partially overlapping bed positions. For all methods, these bed positions were reconstructed jointly in a single combined-bed image space, with separate forward- and back-projections for each bed position and a shared image-domain prior over the combined volume. We adopted this formulation because post-reconstruction sensitivity-weighted stitching can introduce overlap-region artefacts when edge-preserving priors are used \cite{porter_simultaneous_2025}. Bed positions overlapped by $11$ voxels. Phantom data were reconstructed using a single bed position. Projection data are partitioned into staggered (view-interleaved) subsets: $S=9$ for SHKEM (following the comparator study) and, for SVRG-based reconstructions, $18$ subsets per PET bed position for clinical data ($36$ PET subsets in total) and $18$ PET subsets for phantom data; SPECT used $12$ subsets.

\subsubsection{SPECT}
\label{subsubsec:spect_forward}
For SPECT we use a forward model comprising geometric projection with an analytic collimator--detector response and an additive correction obtained via simulation (see below). Images are discretised on a $128\times 128\times 128$ grid with isotropic voxel size $4.18~\mathrm{mm}$. Projection data are partitioned into staggered subsets: $S=12$ for SHKEM and $S=12$ for SVRG-based reconstructions.

The collimator--detector response is represented by a depth-dependent Gaussian with linearly varying width, obtained by fitting an analytic collimator model~\cite{angerSensitivityResolutionLinearity1966}. Bremsstrahlung blur is modelled as an isotropic Gaussian with energy-dependent width $s_{\mathrm{brem}}(E)$ fitted to Y-90 range simulations~\cite{raultFastSimulationYttrium902010}; when multiple energy windows are used, effective blur parameters are obtained by count-weighted averaging across windows.

Scatter and penetration are included via an additive term estimated by an outer reconstruct--simulate loop: we reconstruct, simulate the additive contribution with SIMIND~\cite{ljungbergSIMINDMonteCarlo2015}, hold it fixed for the next reconstruction, and repeat for $10$ outer iterations~\cite{rongDevelopmentEvaluationImproved2012}. This is then fixed for the stochastic optimisation. We initialised the additive term using a learned estimate from a neural network~\cite{xiangDeepNeuralNetwork2020}, trained on data acquired with different scanner, collimator, and energy window settings. To reduce oscillations in the scatter estimation across outer iterations, we used a damped update
\begin{equation}
b^{(k+1)} = \nu\,\hat b^{(k+1)} + (1-\nu)\,b^{(k)},
\end{equation}
with $\nu = 0.2$, where $\hat b^{(k+1)}$ denotes the newly simulated additive estimate and $b^{(k)}$ the previous estimate.

\subsubsection{Co-registration and Warp Operators}
\label{subsubsec:warps}
CTAC provides the anatomical reference for co-registration. In this work, the shared reconstruction grid introduced in Section~\ref{sec:recon_basics} is chosen to be the PET grid. The PET CTAC is therefore downsampled to the PET reconstruction grid and used both to define the CT-derived attenuation map $\mu$ and as the anatomical reference space for the joint reconstruction, so that $W_{\mathrm{PET}}$ reduces to the identity on the shared grid.

The PET and SPECT CTAC volumes were first retained in their native acquisition grids. Since patient arm positioning differed between modalities (arms down in PET, arms up in SPECT), we restricted the registration to the trunk. To do this, a trunk mask was obtained automatically from each CTAC using TotalSegmentator \cite{wasserthal_totalsegmentator_2023}. The native-grid SPECT CTAC was then registered to the PET CTAC by first estimating a rigid transformation and subsequently refining it with a deformable registration initialised from the rigid solution. The resulting native-SPECT-to-PET deformation defines $W_{\mathrm{SPECT}}$.

All coupled-prior and optimisation variables are represented on the PET/shared grid, so SPECT images are mapped into PET space via $W_{\mathrm{SPECT}}$. In the data-fidelity term, the corresponding operator $W_{\mathrm{SPECT}}^\ast$ denotes the discrete adjoint of this resampling operator, not a matrix transpose. Linear interpolation was used for the resampling operators so that the forward and adjoint warp pair were implemented consistently in the optimisation.


\subsection{Image analysis}
\label{subsec:image_analysis}

All image analysis was performed on PET reconstructions to enable direct comparison with Deidda \textit{et al.} \cite{deiddaTripleModalityImage2023}, where PET is the modality of interest and SPECT provides guidance only. SPECT reconstructions are shown for completeness but were not included in the quantitative evaluation. Reconstruction hyperparameters were tuned once using the clinical ``training patient" (SIRT3) and then fixed for all phantom and clinical reconstructions. This single-patient tuning strategy was chosen to avoid repeated adjustment on the evaluation cohort, but it does not constitute a robustness study of the selected weights. Throughout, methods are compared against the reference \wdtnv, with \dtv, \wtnv, and SHKEM as comparators.

\subsubsection{Phantom data}

We assessed voxel-wise repeatability of PET across bootstrap noise realisations and PET quantification via a recovery coefficient (RC) relative to the known true sphere-to-background ratio.

\begin{figure}[htbp]
    \centering
    \begin{subfigure}[b]{0.39\linewidth}
        \centering
        \includegraphics[width=\linewidth,trim=0 0 0 1.8in,clip]{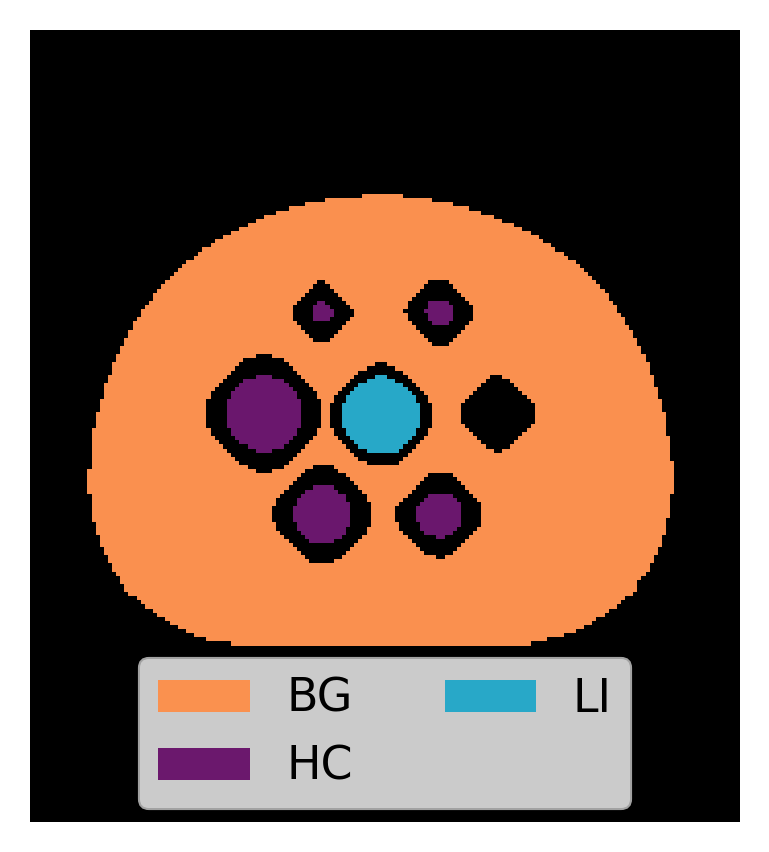}
        \caption*{}
        \label{fig:bootstrap_mask}
    \end{subfigure}%
    \hfill
    \begin{subfigure}[b]{0.58\linewidth}
        \centering
        \includegraphics[width=\linewidth,trim=0 0 0 1.8in,clip]{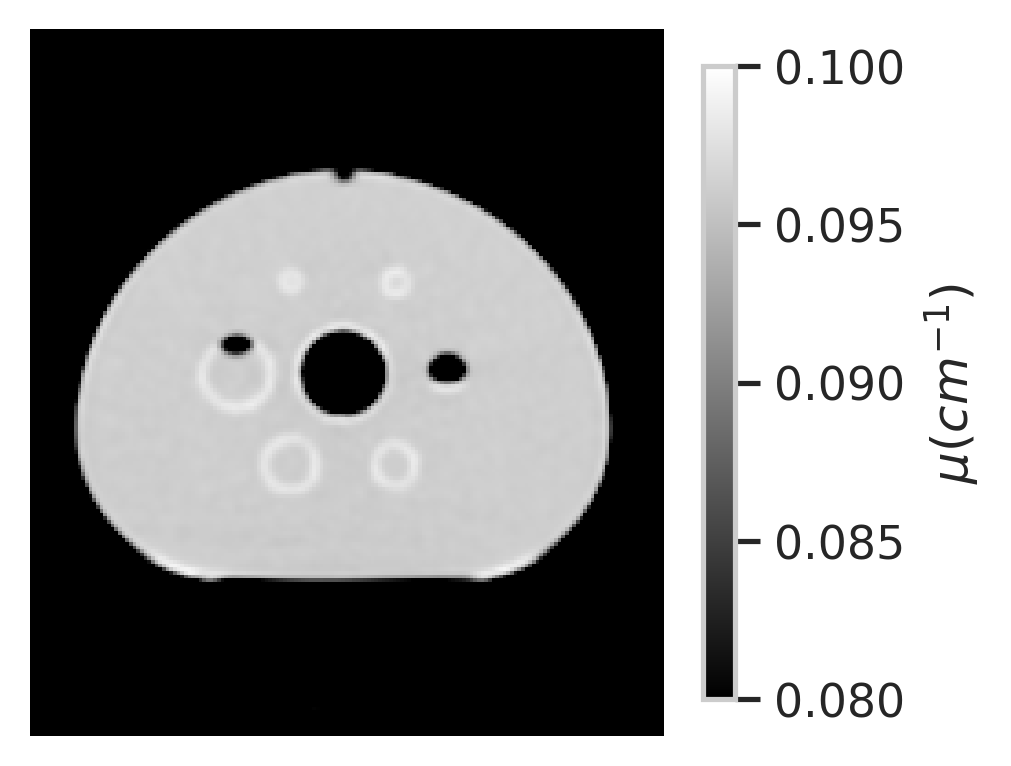}
        \caption*{}
        \label{fig:bootstrap_umap}
    \end{subfigure}
    \caption{Left: VOIs (warm background (BG), lung insert (LI), and high-count (HC) VOIs). Right: attenuation map showing insufficient filling, sphere edges, and the lung insert.}
    \label{fig:nema_bootstrap_overview}
\end{figure}

Three volumes of interest (VOI) were defined (Fig.~\ref{fig:nema_bootstrap_overview}): (1) a high-count (HC) VOI comprising five filled spheres (one sphere was excluded due to filling issues), (2) a warm-background (BG) VOI comprising the phantom volume excluding the spheres plus a \SI{10}{mm} margin and excluding the lung insert plus a \SI{5}{mm} margin, and (3) the cold lung insert (LI).

Let $x^{(r)}_{b,j}$ denote the reconstructed PET intensity at voxel $j$ for method $r$ in bootstrap realisation $b\in\{1,\ldots,B\}$. For each method and voxel, we computed the inter-realisation mean, $\mu^{(r)}_j$, and standard deviation, $s^{(r)}_j$, defined as
\begin{equation}
    \mu^{(r)}_j := \frac{1}{B}\sum_{b=1}^{B} x^{(r)}_{b,j}, \quad
    s^{(r)}_j := \sqrt{\frac{\sum_{b=1}^{B}\left(x^{(r)}_{b,j}-\mu^{(r)}_j\right)^2}{B-1}}.
\end{equation}
Voxel-wise repeatability across bootstrap realisations was quantified by the coefficient of variation
\begin{equation}
\mathrm{CoV}^{(r)}_j = \frac{s^{(r)}_j}{\mu^{(r)}_j}.
\end{equation}
For a VOI $\mathcal{V}$, repeatability was summarised by the VOI-averaged CoV
\begin{equation}
T^{(r)}_{\mathcal{V}}=\frac{1}{|\mathcal{V}|}\sum_{j\in \mathcal{V}} \mathrm{CoV}^{(r)}_{j}.
\end{equation}
To assess statistical significance of inter-bootstrap CoV, we compared each method $r$ with a baseline $r_0$ (\wdtnv) using a paired row-swap permutation test on $\log(T^{(r)}_{\mathcal{V}})$. Let $X^{(r)}_{\mathcal{V}}\in\mathbb{R}^{B\times|\mathcal{V}|}$ be the bootstrap-by-voxel stack restricted to VOI $\mathcal{V}$. The observed difference is $\Delta = \log(T^{(r_0)}_{\mathcal{V}}) - \log(T^{(r)}_{\mathcal{V}})$. Under the null of exchangeability, each bootstrap index is independently swapped between the paired stacks with probability $1/2$ to form permuted stacks $\tilde{X}^{(r_0)}_{\mathcal{V}}$, $\tilde{X}^{(r)}_{\mathcal{V}}$ and permuted differences $\Delta_{\mathrm{perm}}$. Using $N_{\mathrm{perm}}=5000$ permutations, the two-sided p-value is
\begin{equation}
p = \frac{1 + \sum_{i=1}^{N_{\mathrm{perm}}} \mathbf{1}\!\left(|\Delta_{\mathrm{perm},i}|\ge |\Delta|\right)}{N_{\mathrm{perm}}+1}.
\end{equation}
where $\mathbf{1}(\cdot)$ denotes the indicator function, equal to $1$ if its argument is true and $0$ otherwise. P-values were Benjamini–Hochberg (BH) adjusted across methods within each VOI \cite{benjaminiControllingFalseDiscovery1995}. No thresholding was applied to exclude near-zero voxels, so CoV estimates in the cold lung insert should be interpreted as a noise-realisation stability measure rather than a conventional percentage variability around a substantial mean signal.

Quantification was assessed using a recovery coefficient (RC) for spheres with diameters \SI{10}{mm}, \SI{13}{mm}, \SI{22}{mm}, \SI{28}{mm}, and \SI{37}{mm}. Let $\bar{C}^{(r)}_{b,s}$ denote the mean reconstructed intensity within sphere $s$ for method $r$ and bootstrap $b$, and let $\bar{C}^{(r)}_{b,\mathrm{bkg}}$ denote the mean within the warm-background VOI for the same $r,b$. With known true sphere-to-background ratio $R_{\mathrm{true}}=8$, we defined
\begin{equation}
\mathrm{RC}^{(r)}_{b,s} = \frac{\bar{C}^{(r)}_{b,s}}{\bar{C}^{(r)}_{b,\mathrm{bkg}}\,R_{\mathrm{true}}}.
\end{equation}

We treated bootstrap realisations as experimental units and modelled log recovery coefficients with a linear mixed model. Let $r_0=\wdtnv$ again denote the reference method and define
\begin{equation}
y_{b,r,s} := \log\!\left(\mathrm{RC}^{(r)}_{b,s}\right), \qquad b=1,\ldots,B .
\end{equation}
We fit the mixed model
\begin{align}
y_{b,r,s} &= \beta_0 + \alpha_r + \gamma_s + (\alpha\gamma)_{r,s} + u_b + \varepsilon_{b,r,s}, \nonumber \\
u_b &\overset{\mathrm{iid}}{\sim} \mathcal{N}(0,\tau_b^2),\qquad
\varepsilon_{b,r,s} \overset{\mathrm{iid}}{\sim} \mathcal{N}(0,\tau^2),
\end{align}
where $\alpha_r$ is the fixed effect of reconstruction method (with reference $r_0=\wdtnv$ under treatment coding), $\gamma_s$ is the fixed effect of sphere (size), $(\alpha\gamma)_{r,s}$ is the method-by-sphere interaction fixed effect, and $u_b$ is a random intercept for bootstrap replicate $b$. Parameters were estimated by restricted maximum likelihood (REML).

For each sphere $s$ and comparator $r\neq r_0$, the log-scale contrast of interest is
\begin{equation}
\Delta_{r,s} := \mathbb{E}\!\left[y_{b,r,s}-y_{b,r_0,s}\right]
            = \mathbf{L}_{r,s}^\top \boldsymbol{\beta},
\end{equation}
where $\boldsymbol{\beta}$ is the fixed-effect coefficient vector and $\mathbf{L}_{r,s}$ is the corresponding contrast vector. We estimate $\Delta_{r,s}$ by $\widehat{\Delta}_{r,s}=\mathbf{L}_{r,s}^\top \widehat{\boldsymbol{\beta}}$.

For each sphere size, this contrast tests whether the mean log-RC differs between comparator $r$ and \wdtnv. Because the model is fitted on the log scale, $\Delta_{r,s}$ represents the log ratio of mean RC for comparator $r$ relative to \wdtnv for sphere $s$; after exponentiation, $\exp(\Delta_{r,s})<1$ indicates lower recovery than \wdtnv. Inference for these pre-specified fixed-effect contrasts used standard Wald tests for linear mixed-effects models~\cite{laird_random-effects_1982, pinheiro_fitting_2000-1}:
\begin{equation}
z_{r,s} =
\frac{\widehat{\Delta}_{r,s}}
{\mathrm{SE}(\widehat{\Delta}_{r,s})},
\qquad
p_{r,s}=2\left(1-\Phi\!\left(|z_{r,s}|\right)\right),
\end{equation}
where $\Phi(\cdot)$ is the standard normal cumulative distribution function. False discovery rate was controlled at $\alpha=0.05$ using the Benjamini--Hochberg procedure~\cite{benjaminiControllingFalseDiscovery1995} across the planned many-to-one method comparisons within each sphere.

\subsubsection{Clinical data}

\begin{figure}[htbp]
    \centering
    \begin{subfigure}[b]{0.32\linewidth}
        \centering
        \includegraphics[width=\linewidth]{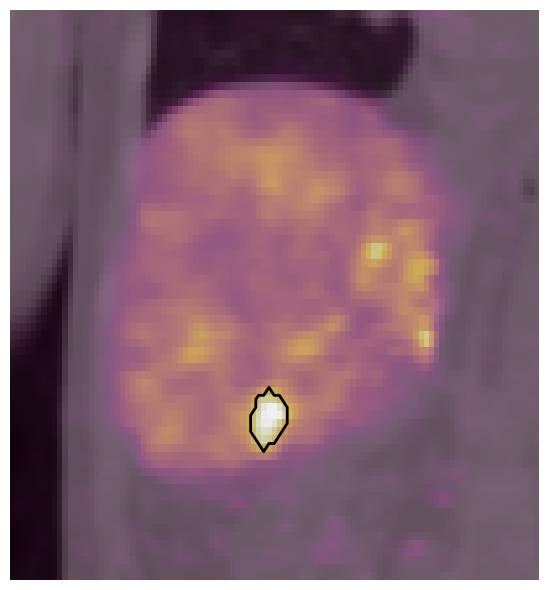}
        \caption*{}
        \label{fig:sirt3_lesion1}
    \end{subfigure}%
    \hfill
    \begin{subfigure}[b]{0.32\linewidth}
        \centering
        \includegraphics[width=\linewidth]{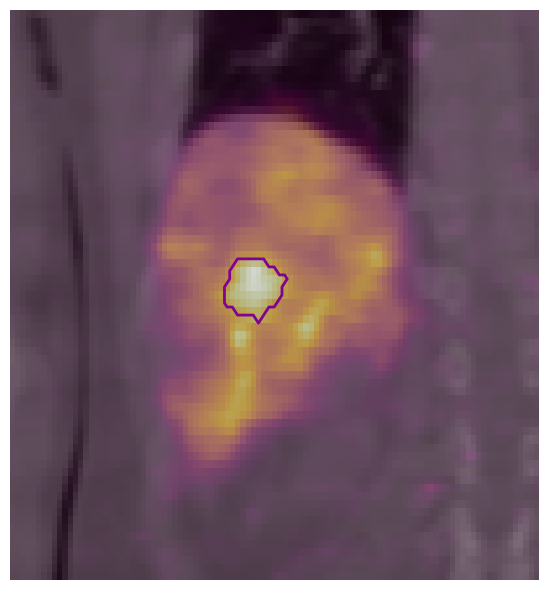}
        \caption*{}
        \label{fig:sirt3_lesion2}
    \end{subfigure}%
    \hfill
    \begin{subfigure}[b]{0.32\linewidth}
        \centering
        \includegraphics[width=\linewidth]{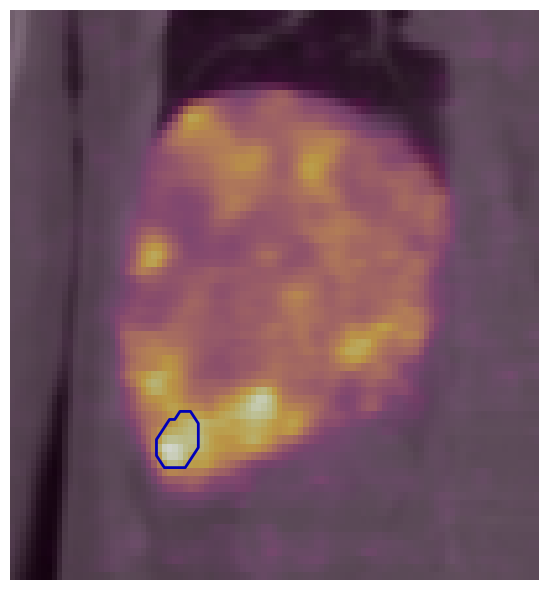}
        \caption*{}
        \label{fig:sirt3_lesion3}
    \end{subfigure}%
    \vspace{-0.4cm}
    \begin{subfigure}[b]{0.32\linewidth}
        \centering
        \includegraphics[width=\linewidth]{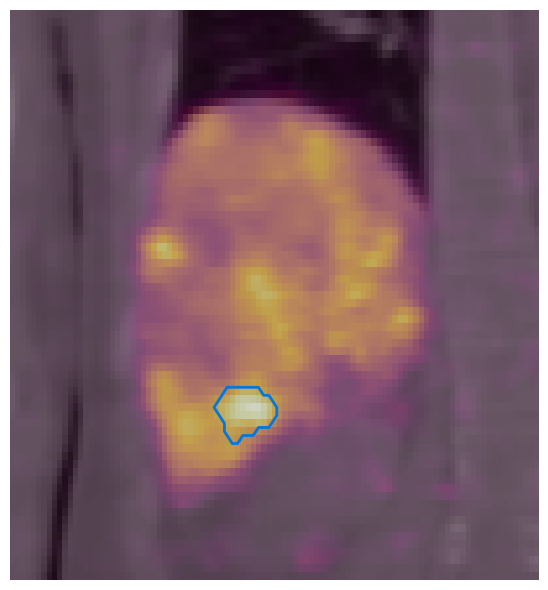}
        \caption*{}
        \label{fig:sirt3_lesion4}
    \end{subfigure}
    \hfill
    \begin{subfigure}[b]{0.32\linewidth}
        \centering
        \includegraphics[width=\linewidth]{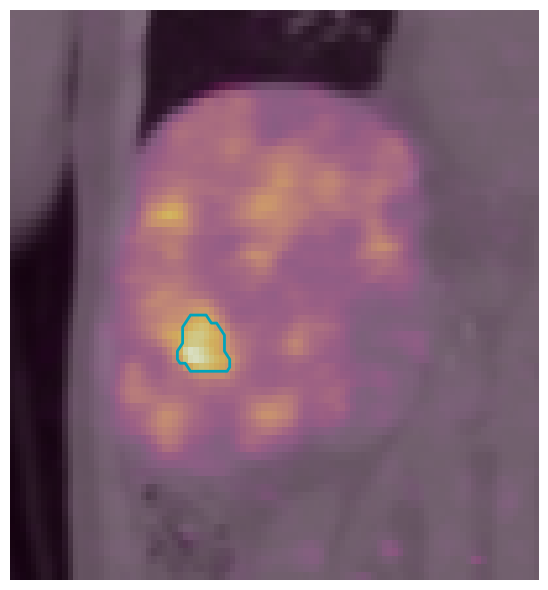}
        \caption*{}
        \label{fig:sirt3_lesion5}
    \end{subfigure}%
    \hfill
    \begin{subfigure}[b]{0.32\linewidth}
        \centering
        \includegraphics[width=\linewidth]{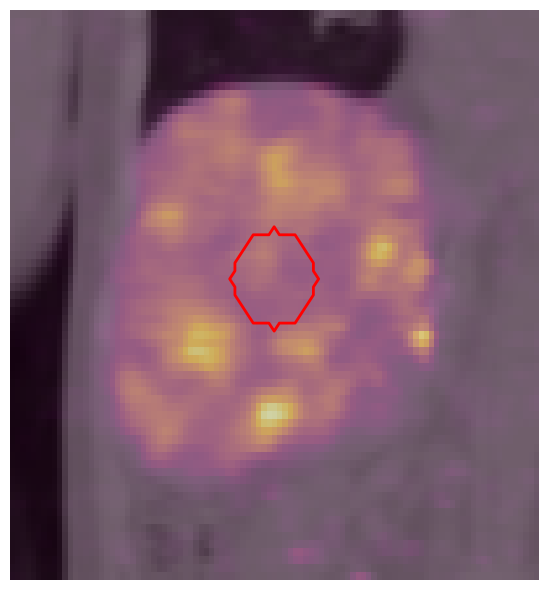}
        \caption*{}
        \label{fig:sirt3_background}
    \end{subfigure}
    \caption{Visualisation of the five identified lesions and the background reference region for patient SIRT3. Shown on Q.Clear vendor reconstructions (with registered attenuation maps). Top row: lesions 1--3. Bottom row: lesions 4--5 and background.}
    \label{fig:sirt3_rois}
\end{figure}

Tumour VOIs were delineated once per patient on the vendor (Q.Clear, $\beta$-4000) PET reconstruction using an automatic gradient-based method \cite{mikellImpact90YPET2018} and propagated unchanged to all reconstructions so that differences reflected reconstruction effects rather than VOI definition. A background VOI was drawn manually in homogeneous, non-lesional liver parenchyma, avoiding motion and visible artefacts (Fig.~\ref{fig:sirt3_rois}).

For patient $p$, lesion $l$, and method $r$, the tumour-to-background ratio (TBR) was
\begin{equation}
\mathrm{TBR}_{p,l,r}=\frac{\bar{C}_{p,l,r}}{\bar{C}_{p,\mathrm{bkg},r}},
\end{equation}
where $\bar{C}_{p,l,r}$ is the lesion mean and $\bar{C}_{p,\mathrm{bkg},r}$ is the background mean. Background noise was summarised per patient and method by the background coefficient of variation (reported as a percentage)
\begin{equation}
\mathrm{CoV}_{p,r}(\%) = 100\,\frac{s_{p,r}}{\bar{C}_{p,\mathrm{bkg},r}},
\end{equation}
where $s_{p,r}$ is the within-background-VOI standard deviation.

To account for multiple lesions per patient and repeated measurements across methods, we analysed the log TBR using the linear mixed-effects model
\begin{equation}
\begin{aligned}
y_{p,l,r} &:= \log\!\left(\mathrm{TBR}_{p,l,r}\right)
= \beta_0 + \alpha_r + u_p + v_{p,l} + \varepsilon_{p,l,r},\\
u_p &\overset{\mathrm{iid}}{\sim} \mathcal{N}(0,\tau^2_{\mathrm{p}}),\quad
v_{p,l} \overset{\mathrm{iid}}{\sim} \mathcal{N}(0,\tau^2_{\mathrm{l}}),\quad
\varepsilon_{p,l,r} \overset{\mathrm{iid}}{\sim} \mathcal{N}(0,\tau^2),
\end{aligned}
\end{equation}
where $\alpha_r$ is the fixed effect of method (reference $r_0=\wdtnv$), $u_p$ is a patient random intercept, and $v_{p,l}$ is a lesion-within-patient random intercept. Parameters were estimated by REML.

For each comparator method $r\neq r_0$, inference targeted the many-to-one contrast on the log scale
\begin{equation}
\Delta_r := \alpha_r - \alpha_{r_0}.
\end{equation}
This contrast represents the log TBR ratio for comparator $r$ relative to \wdtnv. Results are reported on the original scale as $\exp(\Delta_r)$, with values below one indicating lower TBR than \wdtnv. The $95\%$ confidence interval was obtained by exponentiating the corresponding Wald interval for $\Delta_r$, and two-sided $p$-values were obtained from standard Wald tests for these pre-specified contrasts~\cite{laird_random-effects_1982, pinheiro_fitting_2000-1}.

Background noise was analysed analogously at the patient level using
\begin{equation}
\begin{aligned}
z_{p,r} &:= \log\!\left(\mathrm{CoV}_{p,r}(\%)\right)
= \beta_0 + \alpha_r + u_p + \varepsilon_{p,r},\\
u_p &\overset{\mathrm{iid}}{\sim} \mathcal{N}(0,\tau_{\mathrm{p,CoV}}^2),\qquad
\varepsilon_{p,r} \overset{\mathrm{iid}}{\sim} \mathcal{N}(0,\psi^2),
\end{aligned}
\end{equation}
where $\alpha_r$ is the fixed effect of method (reference $r_0=\wdtnv$ under treatment coding) and $u_p$ is a patient-level random intercept. Parameters were estimated by REML. Method effects were summarised as CoV ratios $\exp(\alpha_r-\alpha_{r_0})$, where values below one indicate lower background CoV than \wdtnv. Wald $95\%$ confidence intervals and two-sided $p$-values were computed analogously to the TBR analysis.

False discovery rate across the three planned comparisons within each endpoint (TBR and background CoV) was controlled at $\alpha=0.05$ using the Benjamini--Hochberg procedure~\cite{benjaminiControllingFalseDiscovery1995}. BH-adjusted $p$-values are reported with thresholds $p<0.05$ (*), $p<0.01$ (**), and $p<0.001$ (***).

\subsection{Implementation Details}
\label{subsec:implementation}
All reconstructions were implemented in the Synergistic Reconstruction Framework (SIRF)\cite{ovtchinnikovSIRFSynergisticImage2017} using the Software for Tomographic Image Reconstruction (STIR) projector backends \cite{thielemansSTIRSoftwareTomographic2012, schrammPARALLELPROJOpensourceFramework2024, fusterIntegrationAdvanced3D2013} and the Core Imaging Library \cite{jorgensenCoreImagingLibrary2021} with custom code for the proposed regulariser. We used the same SHKEM implementation as in~\cite{deiddaTripleModalityImage2023}, extended here to the same joint multi-bed reconstruction setting used for the variational methods on clinical PET data.

\section{Results}
\label{sec:results}
Parameter sweeps over $\omega_{\mathrm{PET}}$ and $\omega_{\mathrm{SPECT}}$ (and over $\omega_{\mathrm{PET}}$ for \dtv) are shown in Fig.~\ref{fig:tradeoff}. For each method, we selected a value on the Pareto frontier \cite{paretoNewTheoriesEconomics1897} of the TBR--$\mathrm{CoV}_{\mathrm{bkg}}$ trade-off, at an operating point with $\mathrm{CoV}_{\mathrm{bkg}}$ approximately matched to the vendor reconstruction. The resulting weights are summarised in Table~\ref{tab:hyperparams}. The isotropic TV sweep is included for reference but is not used in subsequent comparisons.

\begin{table}[hbtp]
\centering
\caption{Selected regularisation weights (tuned on SIRT3) used for all phantom and clinical experiments.}
\label{tab:hyperparams}
\small
\begin{tabular}{lcc}
\toprule
\textbf{Method} & $\boldsymbol{\omega_{\mathrm{PET}}}$ & $\boldsymbol{\omega_{\mathrm{SPECT}}}$ \\
\midrule
\wdtnv & 0.01   & 0.02 \\
\wtnv  & 0.0075 & 0.02 \\
\dtv      & 0.02   & -- \\
\bottomrule
\end{tabular}
\end{table}

For SHKEM, hyperparameters were set following the recommendations of Deidda \textit{et al.} \cite{deiddaTripleModalityImage2023}. The number of subiterations (27) was selected to approximately match the $\mathrm{CoV}_{\mathrm{bkg}}$ of the vendor reconstruction.

\subsection{Phantom Data}
\label{subsec:phantom_results}

\begin{figure}[htbp]
    \centering
    \begin{subfigure}[b]{0.99\linewidth}
        \centering
        \includegraphics[width=\linewidth]{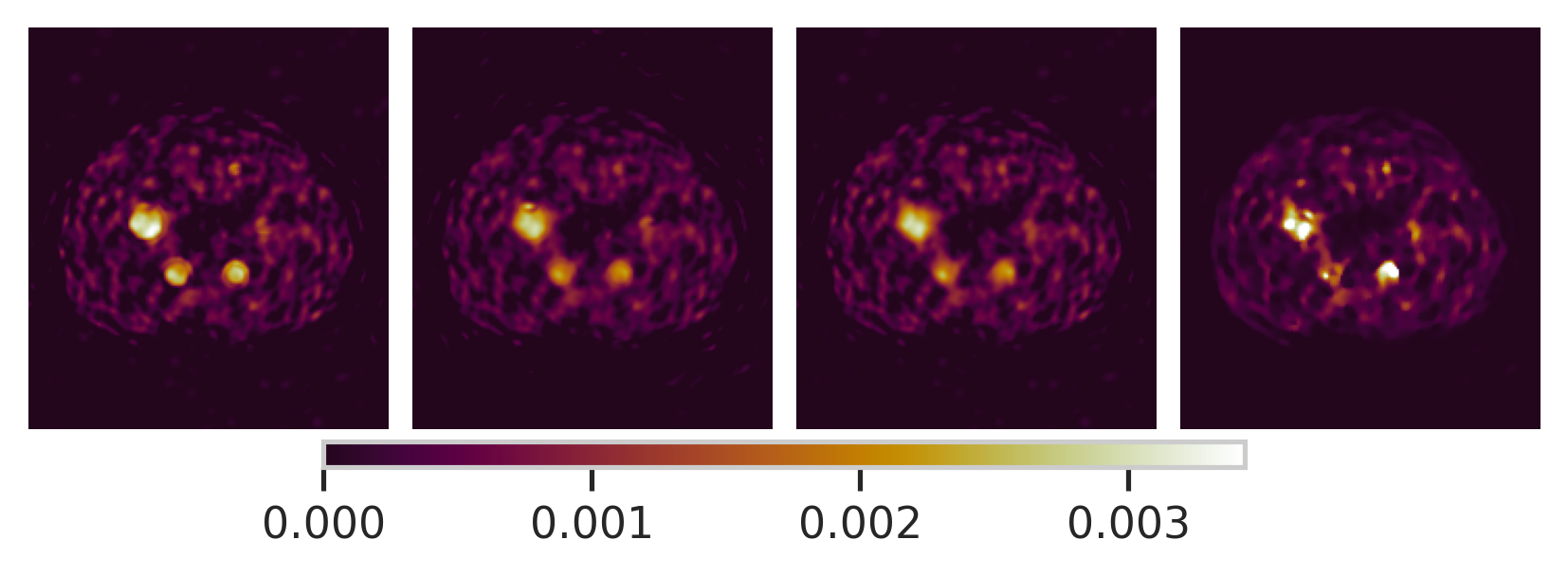}
        \label{fig:phantom_random_pet}
    \end{subfigure}%
    \vspace{-0.5cm}
    \begin{subfigure}[b]{0.99\linewidth}
        \centering
        \includegraphics[width=\linewidth]{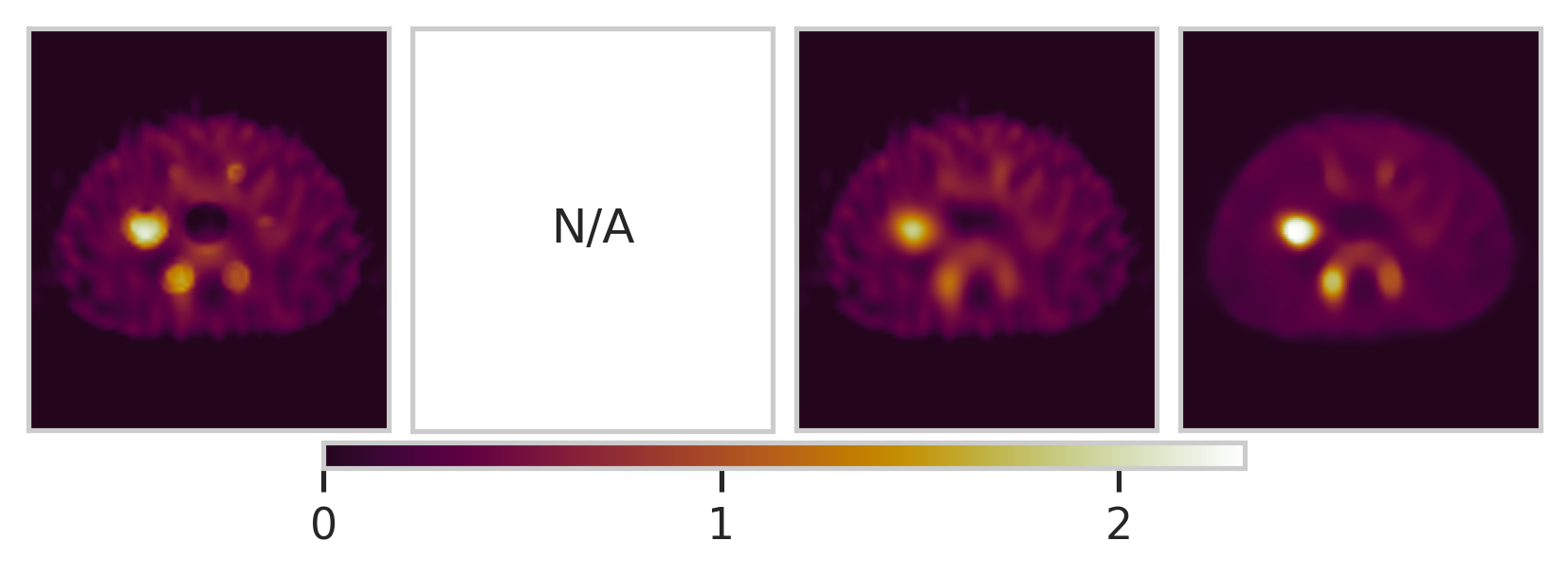}
        \label{fig:phantom_random_spect}
    \end{subfigure}
    \caption{Visual comparison of NEMA phantom reconstructions using different algorithms for a randomly chosen bootstrap. From left to right: \wdtnv, \dtv, \wtnv, and SHKEM (shown at subiteration 27). PET reconstructions are shown on the top row and SPECT on the bottom row (SPECT guidance image for SHKEM). Images are displayed in reconstruction intensity units, proportional to activity concentration.}
    \label{fig:phantom_recon_comparison}
\end{figure}

A representative reconstruction from a single bootstrapped PET noise realisation is shown in Fig.~\ref{fig:phantom_recon_comparison}. We display PET (top row) and the corresponding SPECT image (bottom row); for SHKEM this is the separately reconstructed guidance SPECT, while \dtv does not reconstruct SPECT.

\subsubsection{Quantification}
\begin{figure*}
    \centering
    \includegraphics[width=\linewidth]{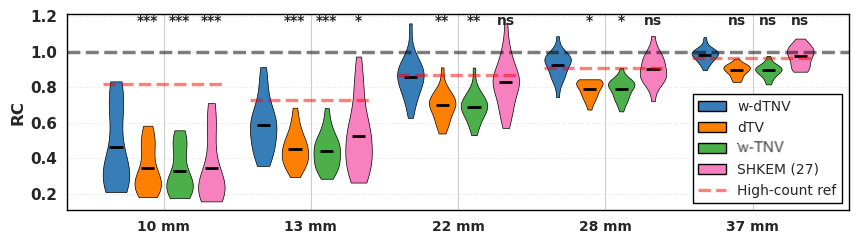}
    
    \caption{Recovery coefficient ($RC$) for NEMA phantom spheres for \wdtnv, \dtv, \wtnv and SHKEM (shown at subiteration 27). Also shown are the equivalent $RC_{s}$ for a high-count, long-OSEM reconstruction (24ss, 240it). The grey line demonstrates perfect recovery. Statistical significance is reported for paired comparisons versus \wdtnv (Benjamini–Hochberg adjusted): $^*$ $p<0.05$, $^{**}$ $p<0.01$, $^{***}$ $p<0.001$.}
    \label{fig:rc_spheres_comparison}
\end{figure*}

Quantitative analysis demonstrated that \wdtnv achieved significantly higher recovery than \dtv and isotropic \wtnv across all sphere sizes $\leq$ \SI{22}{\milli\metre}. \wdtnv showed similar recovery to SHKEM in the larger spheres but significantly improved recovery for the two smallest spheres ($p<0.05$) (Table~\ref{tab:effect_ratio_rc}, Figure~\ref{fig:rc_spheres_comparison}).

\begin{table}[htbp]
\centering
\footnotesize
\setlength{\tabcolsep}{5pt}
\begin{tabular}{l l S[table-format=1.4] l l}
\toprule
Method & $s$ & {$\exp(\Delta_{r,s})$} & {95\% CI} & {$p_{\mathrm{BH}}$} \\
\midrule
\dtv & 37\,mm & 0.9113 & [0.8001,\,1.0380] & 0.2472 \\
       & 28\,mm & 0.8543 & [0.7500,\,0.9730] & 0.0267 \\
       & 22\,mm & 0.8165 & [0.7168,\,0.9300] & 0.0034 \\
       & 13\,mm & 0.7819 & [0.6865,\,0.8906] & 0.00043\\
       & 10\,mm & 0.7706 & [0.6766,\,0.8777] &  $<10^{-4}$ \\
\addlinespace
\wtnv & 37\,mm & 0.9108 & [0.7997,\,1.0374] & 0.2472 \\
        & 28\,mm & 0.8544 & [0.7502,\,0.9732] & 0.0267 \\
        & 22\,mm & 0.8090 & [0.7103,\,0.9215] & 0.0034 \\
        & 13\,mm & 0.7598 & [0.6671,\,0.8654] & 0.0001 \\
        & 10\,mm & 0.7384 & [0.6483,\,0.8410] & $<10^{-4}$ \\
\addlinespace
SHKEM (27) & 37\,mm & 0.9931 & [0.8720,\,1.1312] & 0.9175 \\
           & 28\,mm & 0.9739 & [0.8551,\,1.1093] & 0.6906 \\
           & 22\,mm & 0.9616 & [0.8443,\,1.0953] & 0.5559 \\
           & 13\,mm & 0.8723 & [0.7659,\,0.9936] & 0.0397 \\
           & 10\,mm & 0.7381 & [0.6480,\,0.8407] &  $<10^{-4}$ \\
\bottomrule
\end{tabular}
\caption{Effect ratios $\exp(\Delta_{r,s})$ (comparator vs.\ baseline \wdtnv) with $95\%$ confidence intervals and Benjamini--Hochberg (BH) adjusted $p$-values, reported by sphere size $s$. Values below one indicate lower recovery coefficient than \wdtnv.}
\label{tab:effect_ratio_rc}
\end{table}

\subsubsection{Bootstrap Repeatability}
\begin{figure}[htbp]
    \centering
    \begin{subfigure}[b]{0.99\linewidth}
        \centering
        \includegraphics[width=\linewidth]{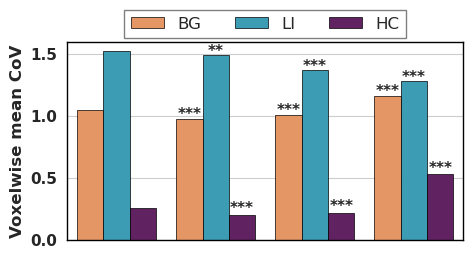}
    \end{subfigure}%
    \caption{Mean voxel-wise coefficient of variation (CoV) across the 20 bootstrapped PET noise realisations, computed within each volume of interest (VOI). Statistical significance is reported for paired comparisons versus \wdtnv (Benjamini–Hochberg adjusted): $^*$ $p<0.05$, $^{**}$ $p<0.01$, $^{***}$ $p<0.001$.}
    \label{fig:phantom_cov}
\end{figure}

Overall, \dtv and \wtnv exhibited the highest voxel-wise repeatability across noise realisations, with the lowest mean voxel-wise CoV in the background and high-count VOIs. \wtnv showed statistically significant differences relative to \wdtnv ($p<0.001$). In contrast, SHKEM showed significantly poorer voxel-wise repeatability in the background and high-count regions ($p<0.05$), but achieved improved repeatability in the low-count lung insert, where it produced the lowest voxel-wise CoV among the methods. Sphere-level CoV trends were consistent with the voxel-wise analysis, with \wtnv and \dtv generally providing the lowest variability across the evaluated sphere diameters and SHKEM the highest. This can be seen visually in Figure~\ref{fig:phantom_cov}.

\subsection{Clinical Data}
\subsubsection{Single-Patient Hyperparameter Optimisation}
\label{subsec:clinical_optimisation}

\begin{figure*}[htbp]
    \centering
    \begin{subfigure}[b]{\textwidth}
        \centering
        \includegraphics[width=\linewidth]{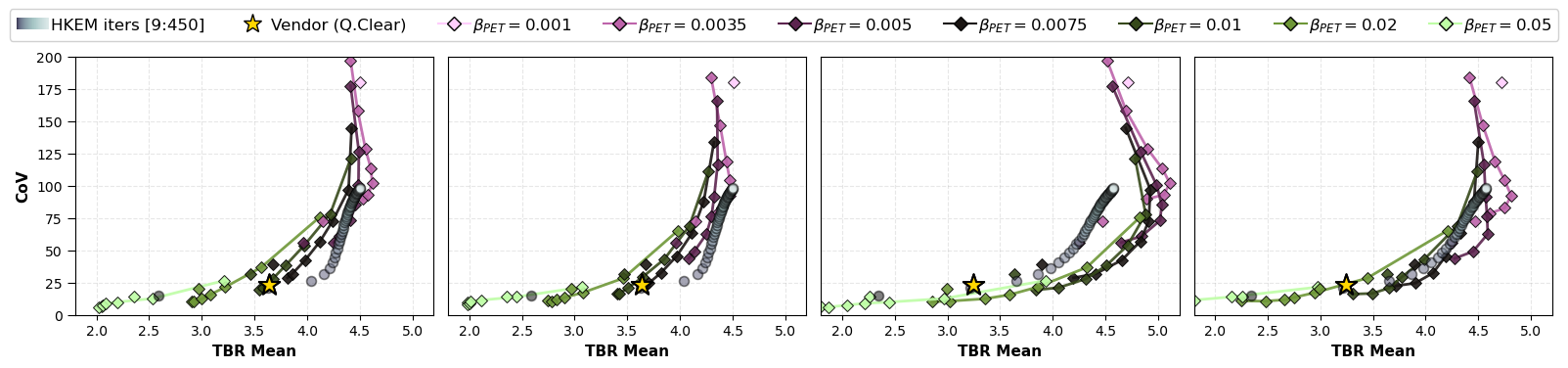}
        \caption*{}
        \label{fig:dtnv_constPET}
    \vspace{-0.5cm}
    \end{subfigure}
    \begin{subfigure}[b]{\textwidth}
        \centering
        \includegraphics[width=\linewidth]{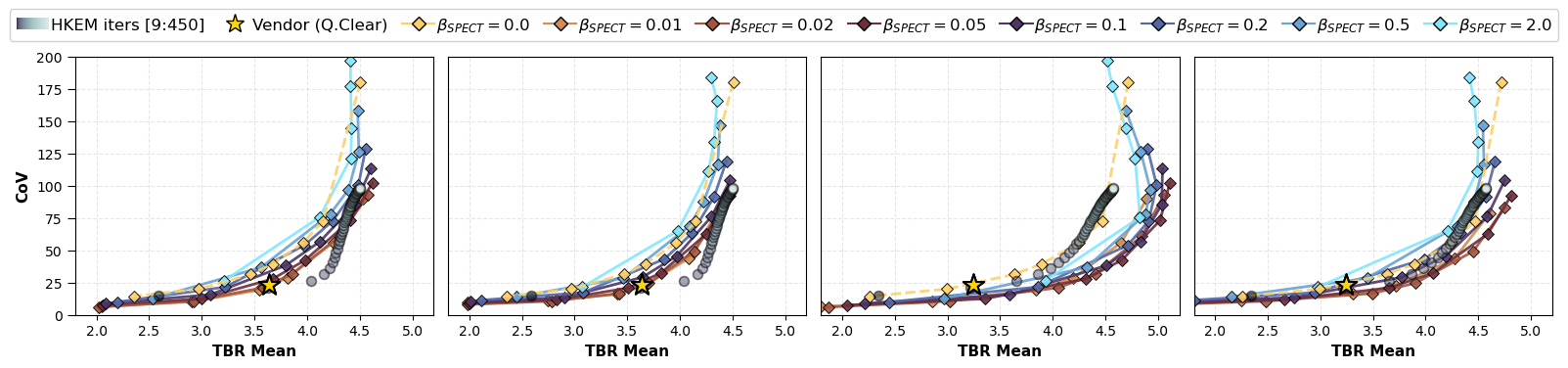}
        \caption*{}
        \label{fig:dtnv_constSPECT}
    \end{subfigure}
    
    \caption{Impact of varying PET weighting ($\omega_{\mathrm{PET}}$) while holding SPECT regularisation constant (top row) and varying SPECT weighting ($\omega_{\mathrm{SPECT}}$) while holding PET regularisation constant (bottom row) on TBR and $\mathrm{CoV}_{\mathrm{bkg}}$ for lesion 1 (first and second columns) and lesion 2 (third and fourth columns). The first and third columns show \wdtnv and the second and fourth show \wtnv.}
    \label{fig:tradeoff}
\end{figure*}

Figure~\ref{fig:tradeoff} shows the tumour-to-background ratio (TBR) versus background coefficient of variation ($\mathrm{CoV}_{\mathrm{bkg}}$) trade-off for two representative lesions. Lesion~1 shows only a modest initial benefit from synergistic information and is largely insensitive to further increases in SPECT weighting; performance remains approximately unchanged over most of the explored range, and degrades only when $\omega_{\mathrm{SPECT}}$ becomes large, consistent with over-regularisation. In contrast, Lesion~2 demonstrates a clear synergistic benefit, exhibiting an improved TBR--$\mathrm{CoV}_{\mathrm{bkg}}$ trade-off.

For Lesion~2, increasing $\omega_{\mathrm{SPECT}}$ improves performance up to a distinct elbow region: the trade-off curves shift towards higher TBR at comparable $\mathrm{CoV}_{\mathrm{bkg}}$ (top row of Figure~\ref{fig:tradeoff}) and the elbow becomes more pronounced (bottom row). Beyond this operating region, further increases in $\omega_{\mathrm{SPECT}}$ reduce TBR and/or increase $\mathrm{CoV}_{\mathrm{bkg}}$, indicating that the synergistic term is being over-weighted. For Lesion~1, by comparison, the corresponding curves show little movement as $\omega_{\mathrm{SPECT}}$ varies, implying limited informative structural content in the SPECT guidance for this lesion, with deterioration only at the highest $\omega_{\mathrm{SPECT}}$ values.

For Lesion~2, the PET image obtained with \wdtnv exhibited features consistent with a CTAC artefact; this is discussed further in Section~\ref{sec:discussion}.

\begin{figure*}[hbtp]
    \centering

    \begin{minipage}[htbp]{0.49\textwidth}
        \centering
        \includegraphics[width=\linewidth]{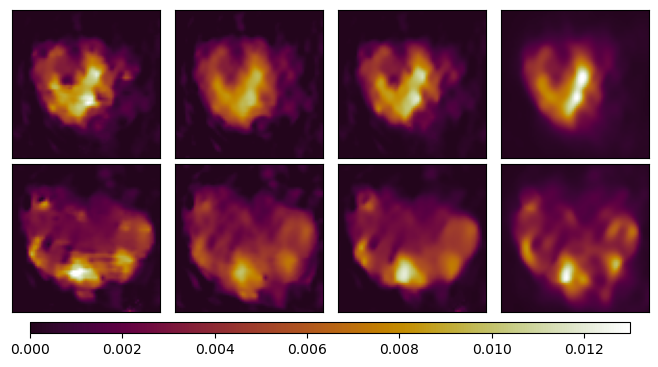}\par
        \includegraphics[width=\linewidth]{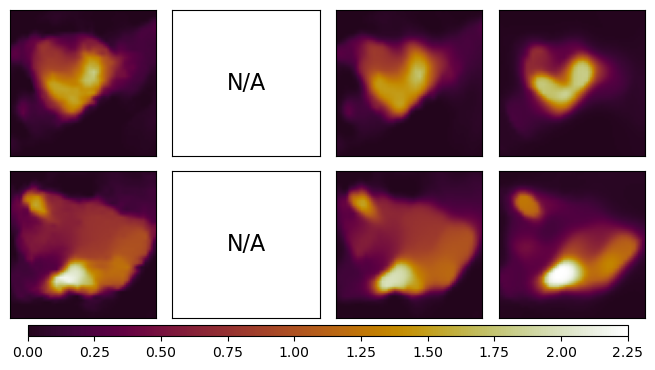}\par
        \textbf{Axial views}.
    \end{minipage}\hfill
    \begin{minipage}[htbp]{0.49\textwidth}
        \centering
        \includegraphics[width=\linewidth]{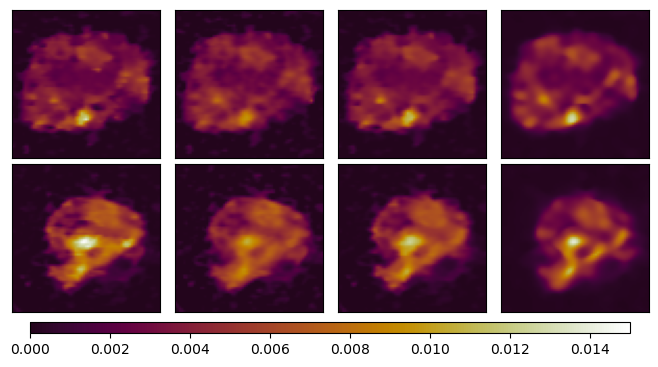}\par
        \includegraphics[width=\linewidth]{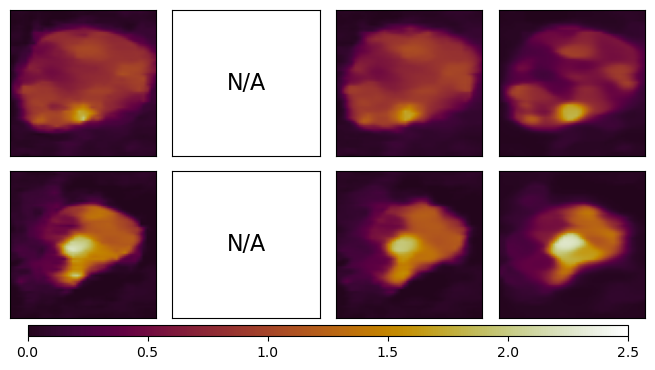}\par
        \textbf{Coronal views}.
    \end{minipage}

    \caption{Visual comparison of reconstruction methods for patient SIRT3 lesion 1 (top row in each quadrant) and lesion 2 (bottom row). Columns show (from left to right) \wdtnv, \dtv, \wtnv and SHKEM (shown at subiteration 27). The upper two quadrants show PET reconstructions, the lower two show SPECT reconstructions or guidance images. Images are displayed in reconstruction intensity units, proportional to activity concentration.}
    \label{fig:sirt3_visuals}
\end{figure*}

Overall, \wdtnv consistently outperformed \wtnv in lesions where synergistic information was informative, while achieving broadly comparable performance in lesions where synergistic information provided limited benefit. Reconstructions centred on lesions 1 and 2 are shown in Figure~\ref{fig:sirt3_visuals}.

\subsubsection{Full Patient Cohort}
\begin{figure}[htbp]
    \centering
    \begin{subfigure}[b]{0.49\linewidth}
        \centering
        \includegraphics[width=\linewidth]{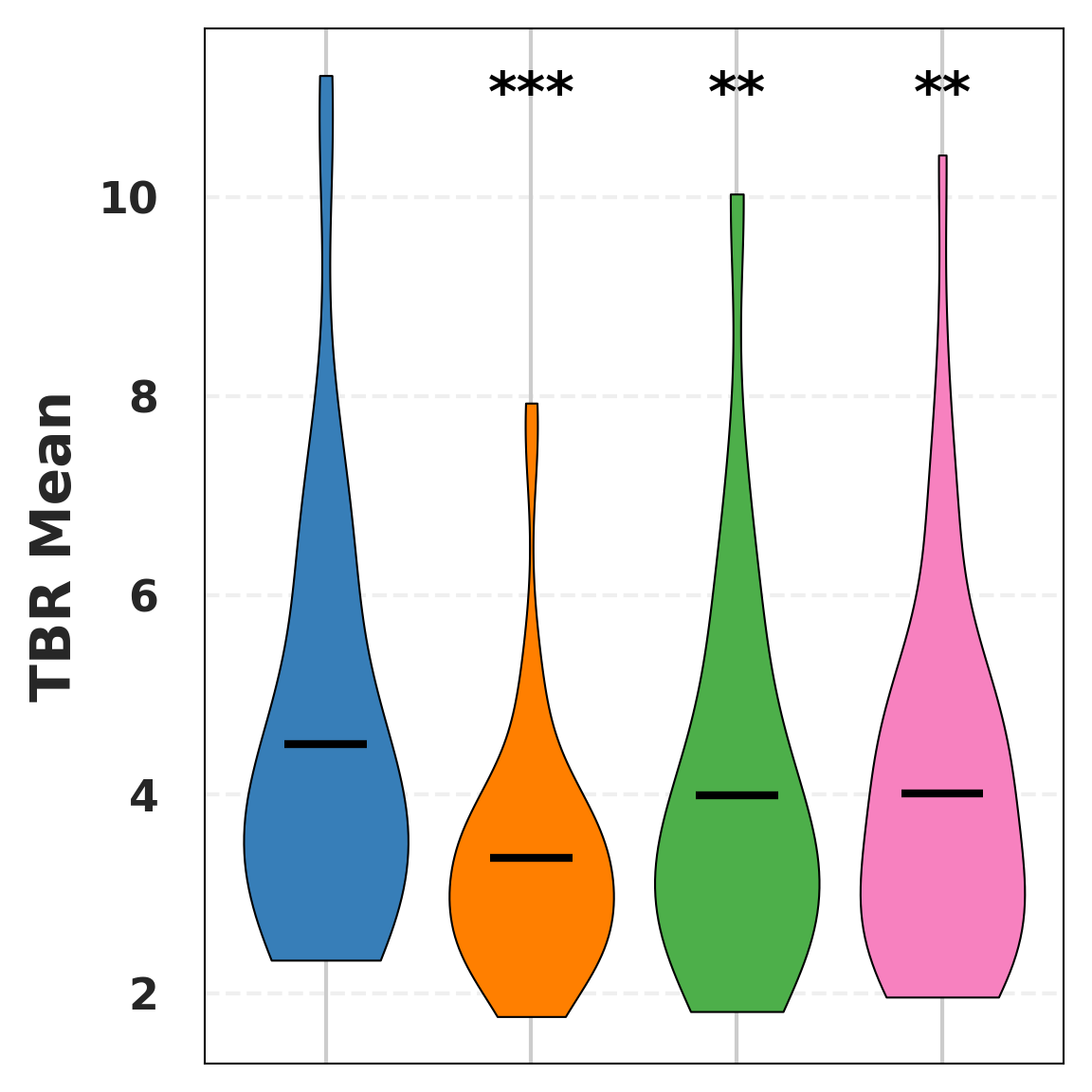}
        \label{fig:patient_tbr}
    \end{subfigure}%
    \hfill
    \begin{subfigure}[b]{0.49\linewidth}
        \centering
        \includegraphics[width=\linewidth]{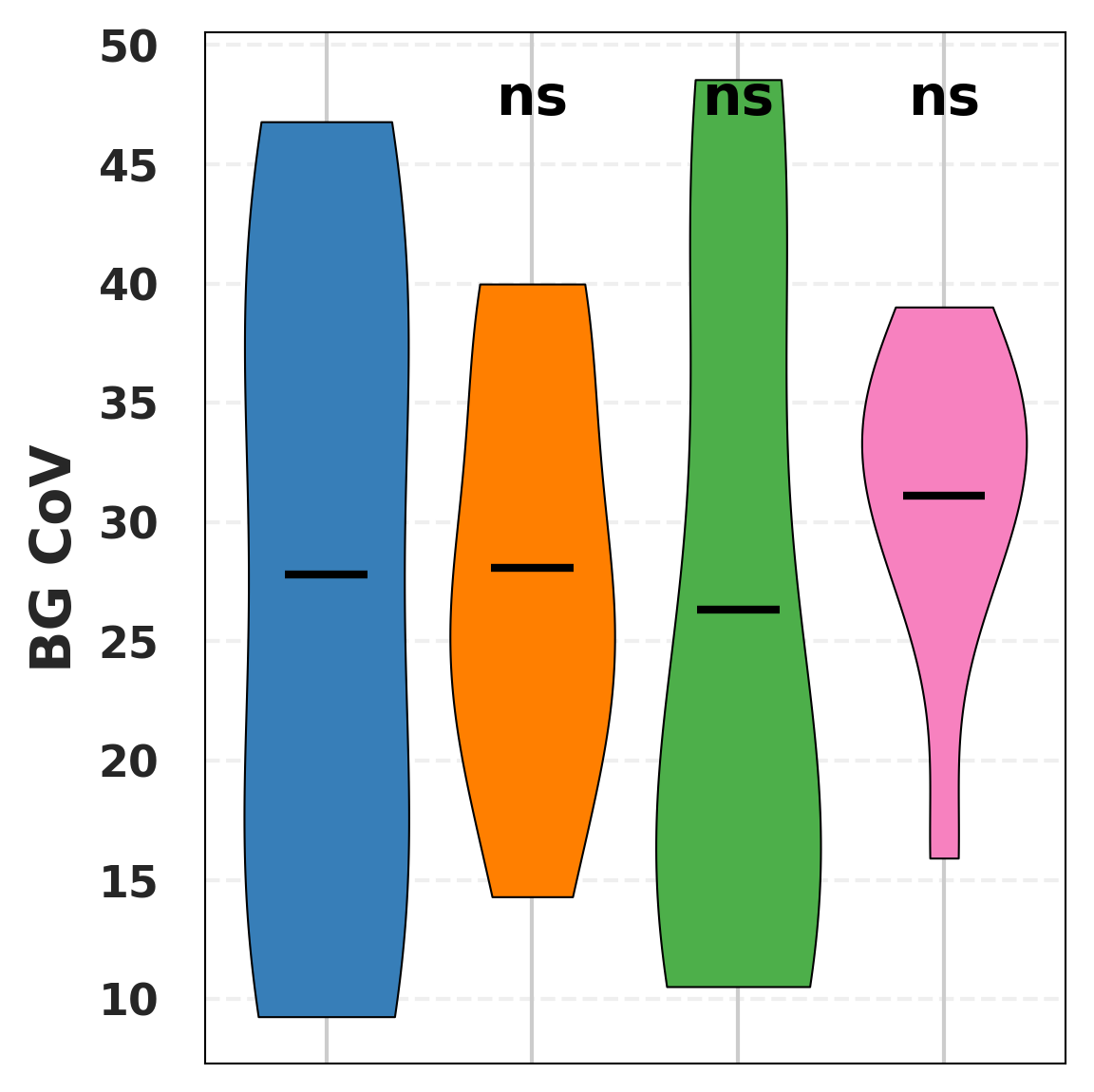}
        \label{fig:patient_cov}
    \end{subfigure}%
    \hfill
    \caption{TBR mean (left) for all lesions and patients, and background CoV (right) across all patients. Blue: \wdtnv, orange: \dtv, green: \wtnv and pink: SHKEM (shown at subiteration 27). Statistical significance is reported for paired comparisons versus \wdtnv (Benjamini–Hochberg adjusted): $^*$ $p<0.05$, $^{**}$ $p<0.01$, $^{***}$ $p<0.001$.}
    \label{fig:patient_summary}
\end{figure}

Overall, \wdtnv delivered improved lesion contrast, achieving significantly higher TBR than \dtv, \wtnv and SHKEM (shown at subiteration 27) (Figure~\ref{fig:patient_summary}, left). SHKEM reconstructions appeared visually smoother and, as also observed in the phantom results, appeared more effective at suppressing cold-background noise.

\begin{table}[htbp]
\centering
\footnotesize
\setlength{\tabcolsep}{5pt}
\begin{tabular}{l S[table-format=1.3] l l}
\toprule
Method & {$\exp(\Delta)$} & {95\% CI} & {$p_{\mathrm{BH}}$} \\
\midrule
\dtv      & 0.802 & [0.758,\,0.849] & $<10^{-4}$ \\
\wtnv     & 0.915 & [0.865,\,0.968] & 0.004 \\
SHKEM (27)  & 0.935 & [0.884,\,0.990] & 0.020 \\
\bottomrule
\end{tabular}
\caption{Effect ratios $\exp(\Delta)$ with $95\%$ confidence intervals and Benjamini--Hochberg (BH) adjusted $p$-values for tumour-to-background ratio (TBR), estimated from the lesion-level mixed-effects model. Values below one indicate lower TBR than \wdtnv.}
\label{tab:effect_ratio_tb}
\end{table}

\begin{table}[htbp]
\centering
\footnotesize
\setlength{\tabcolsep}{5pt}
\begin{tabular}{l S[table-format=1.3] l l}
\toprule
Method & {$\exp(\Delta)$} & {95\% CI} & {$p_{\mathrm{BH}}$} \\
\midrule
\dtv      & 1.183 & [0.903,\,1.549] & 0.35 \\
\wtnv     & 1.007 & [0.833,\,1.167] & 0.984 \\
SHKEM (27)  & 1.339 & [0.933,\,1.913] & 0.093 \\
\bottomrule
\end{tabular}
\caption{Effect ratios $\exp(\Delta)$ with $95\%$ confidence intervals and Benjamini--Hochberg (BH) adjusted $p$-values for background coefficient of variation (CoV), estimated from the patient-level mixed-effects model. Values above one indicate higher background CoV than \wdtnv.}
\label{tab:effect_ratio_cov}
\end{table}

\section{Discussion}
\label{sec:discussion}

In the phantom, \wtnv and \dtv underperformed \wdtnv and SHKEM in recovery, suggesting that functional coupling alone is insufficient and that anatomical guidance (explicit in \wdtnv and implicit via CTAC-informed kernel features in SHKEM) is beneficial for boundary localisation, particularly at smaller sphere sizes. However, this improved recovery came with a repeatability cost: \dtv and \wtnv yielded lower voxel-wise CoV than \wdtnv across VOIs. All three variational methods outperformed SHKEM in background and high-count repeatability, whereas SHKEM was most stable in the cold lung insert. We hypothesise that this is because the SHKEM kernel is partially parameterised by the companion SPECT image, which in the cold insert is itself relatively smooth and less susceptible to salt-and-pepper artefacts and positivity-induced bias that can arise in ultra-low-count PET. Consequently, the kernel weights within the cold region tend to favour local averaging among consistently low-intensity voxels, so that near-zero values are preferentially reinforced. In effect, zeros propagate zeros, yielding increased stability in the cold lung insert despite comparatively weaker repeatability elsewhere.

Notably, \dtv and \wtnv produced broadly similar PET appearances in the phantom, despite \dtv including $\mu$-directional guidance and \wtnv including PET/SPECT coupling. This suggests that, in this experiment, neither CTAC-derived directional guidance alone nor isotropic PET/SPECT coupling alone was sufficient to recover the smallest structures. The improved recovery observed with \wdtnv is therefore consistent with a complementary mechanism: the higher-count SPECT image provides additional functional support for the location of activity transitions, while the $\mu$-derived directional field reduces penalisation of gradients aligned with anatomical boundaries. In this sense, the benefit of \wdtnv appears to arise from combining SPECT-supported functional structure with CTAC-guided edge preservation, rather than from either source of guidance in isolation.

Clinical validation revealed marked heterogeneity in lesion-specific response to directional guidance. Lesion 2 exhibited substantial improvement in the TBR-CoV trade-off with increasing SPECT weighting (Figure~\ref{fig:tradeoff}). Post-hoc examination revealed that this lesion lies adjacent to a region of calcification visible in the \mumap, providing sharp anatomical edges that coincide with the metabolic boundary. Similarly, Lesion 3 demonstrated improved trade-offs under both \wdtnv and \wtnv priors. This lesion abuts the liver on three sides, where the soft-tissue interface generates reasonably strong anatomical contrast. For \wdtnv, the \mumap liver boundary provides explicit directional guidance aligned with the functional uptake margin. For \wtnv, the SPECT boundary may be particularly well-defined at this anatomical location, facilitating effective PET-SPECT coupling even without $\mu$ guidance. This suggests that lesions with anatomically stable boundaries benefit from synergistic reconstruction regardless of the specific prior formulation.

Conversely, Lesion 1 showed much lower benefit from directional guidance and synergistic reconstruction, with trade-off curves remaining static until excessive SPECT weighting introduced negative bias (Figure~\ref{fig:tradeoff}). Visual inspection of the reconstructions and $\mu$ images reveals no obvious anatomical correlate: the PET and SPECT images appear well-aligned, and the \mumap shows no distinctive structural features at the tumour margin. The reason for this lesion's unresponsiveness to synergistic reconstruction remains unclear.

\begin{figure*}[t]
    \centering

    \begin{minipage}[t]{0.49\textwidth}
        \centering
        \includegraphics[width=\linewidth]{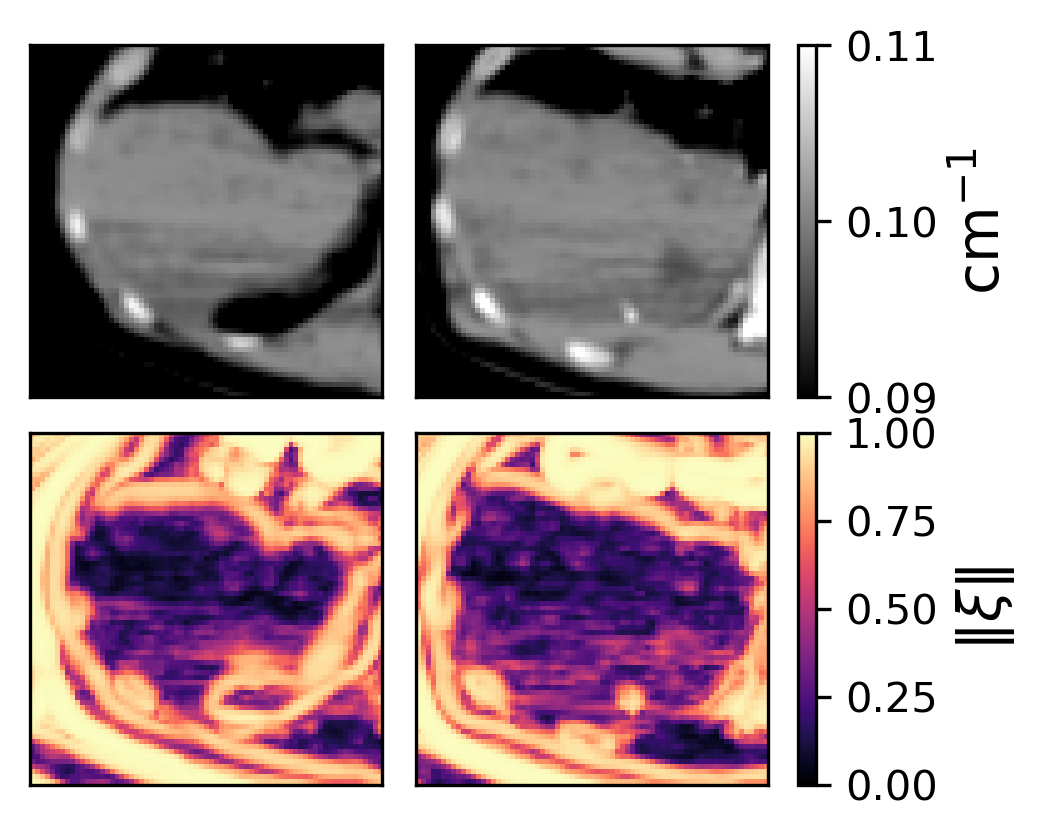}\par
        \vspace{0.0ex}
        \textbf{Axial slice}.              
    \end{minipage}\hfill
    \begin{minipage}[t]{0.49\textwidth}
        \centering
        \includegraphics[width=\linewidth]{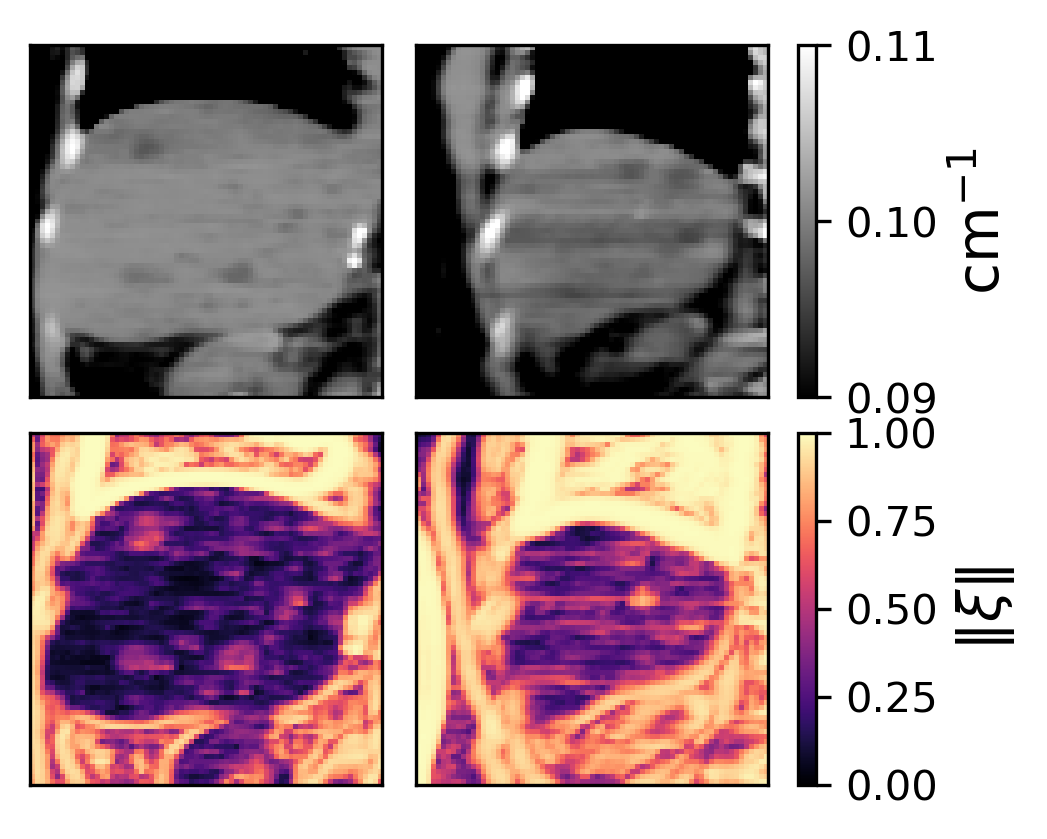}\par
        \vspace{0.0ex}
        \textbf{Coronal slice}.              
    \end{minipage}

    \caption{Patient study. CTAC-derived attenuation map $\mu$ (top row; cm$^{-1}$) and magnitude of the \wdtnv directional weighting field $\|\xi\|$ (bottom row). Columns show Lesion~1 (left) and Lesion~2 (right). CTAC streak / arms-down artefacts are visible in $\mu$ and are reproduced in $\|\xi\|$, providing a pathway for CT artefact transfer into the emission reconstruction; directional-operator smoothing cannot fully compensate.}
    \label{fig:patient_ctac_xi}
\end{figure*}

The clinical study highlights an important practical limitation of \mumap-guided \wdtnv: CTAC artefacts can be transferred into the emission reconstruction through the directional weighting field. In patient SIRT3, the low-dose CTAC was acquired with the arms alongside the torso, which introduced streaking/beam-hardening artefacts in the \mumap. As shown in Figure~\ref{fig:patient_ctac_xi}, these structured \mumap errors are mirrored in the magnitude of the directional weights $\|\xi\|$, implying that the regulariser is being driven by corrupted $\mu$ gradients. The directional-operator smoothing (controlled by $\eta$) can reduce high-frequency CT noise, but it cannot reliably suppress coherent, high-contrast CT artefacts. Consequently, CTAC streaks may imprint into the w-dTNV reconstruction, notably for Lesion~2 in Fig.~\ref{fig:sirt3_visuals}. SHKEM anatomical guidance was derived from the SPECT attenuation map (arms raised), rather than the low-dose PET CTAC used to construct the \mumap-guided directional weights. This difference in the guidance input may contribute to the absence of streak-related artefacts in the SHKEM reconstructions.

\begin{figure}[hbtp]
    \centering
    \includegraphics[width=\linewidth]{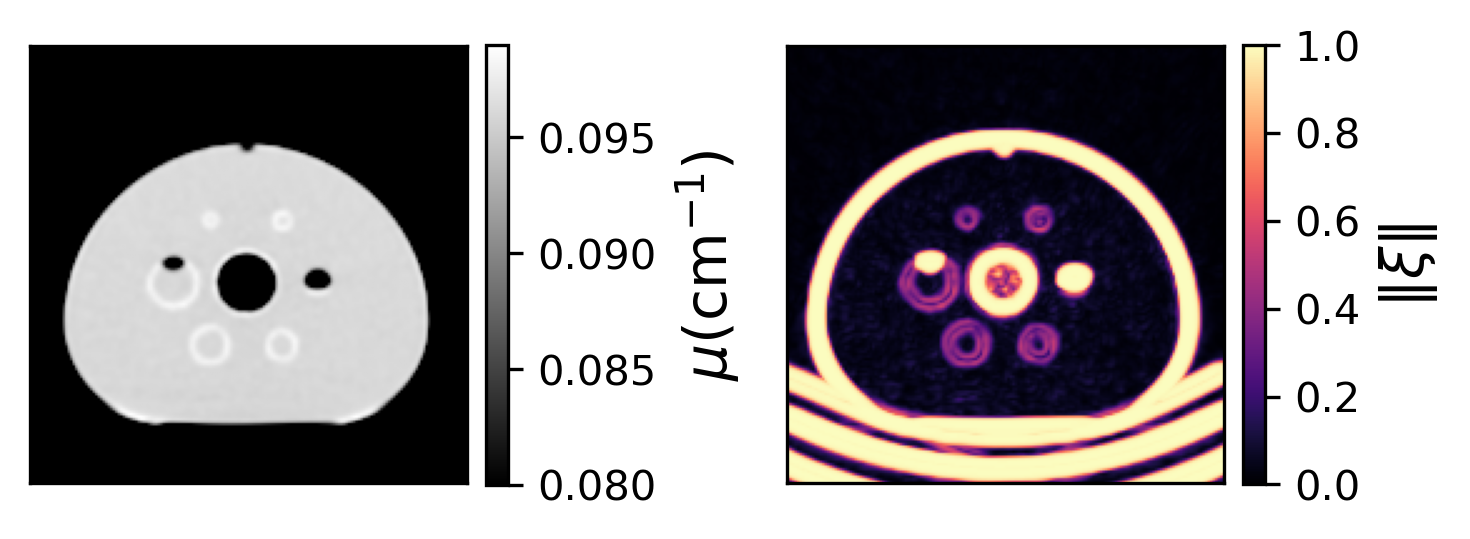}
    \caption{NEMA phantom (central axial slice). CTAC-derived attenuation map $\mu$ (left; cm$^{-1}$) and magnitude of the \wdtnv directional weighting field $\|\xi\|$ (right). With a high-quality CTAC free of prominent streak artefacts, $\|\xi\|$ is well-behaved and concentrates around true boundaries and insert edges, enabling effective \mumap-guided directional smoothing.}
    \label{fig:nema_ctac_xi}
\end{figure}

In contrast, the NEMA phantom CTAC is largely free of such artefacts, and the resulting $\|\xi\|$ follows the true object boundaries and insert edges in a stable manner (Fig.~\ref{fig:nema_ctac_xi}). In this setting, directional smoothing effectively attenuates \mumap-driven high-frequency fluctuations without introducing structured bias. Future work should therefore focus on improving robustness to CTAC artefacts and on characterising the interaction with the smoothing parameter $\eta$. From a practical perspective, if a low-dose CTAC is used to provide directional information, acquisition protocols that minimise artefacts (e.g.\ arms raised where feasible) are strongly recommended. 

To preserve direct comparability with SHKEM, regularisation weights were selected once on the clinical training case and quantitative evaluation focused on PET; the jointly reconstructed SPECT images were assessed qualitatively only. A further limitation is that regularisation weights were selected using a single training patient, so the robustness of these choices across acquisition protocols and lesion presentations was not fully characterised. Although PET/SPECT/CT alignment appeared visually acceptable, the sensitivity of \wdtnv to residual registration error was not independently characterised here and remains an important topic for future work.

Although quantitative analysis in this study focused on PET to enable direct comparison with SHKEM, the jointly reconstructed SPECT images consistently appeared visually less noisy whilst maintaining visual lesion conspicuity. This qualitative behaviour suggests that, beyond using SPECT purely as a source of guiding information for PET, there may be value in treating both reconstructed modalities as clinically relevant outputs of the synergistic framework, particularly in settings where SPECT count statistics are favourable. A rigorous assessment of SPECT quantification and lesion detectability would require modality-specific evaluation protocols (including appropriate lesion delineation) and is therefore left to future work. More broadly, although this work focused on the favourable case in which PET and SPECT image the same underlying $^{90}$Y microsphere distribution, the proposed framework may be extendable to settings with non-identical functional distributions but shared anatomy. Examples include planning--treatment image pairs (e.g.\ $^{99\mathrm{m}}$Tc-MAA versus post-treatment $^{90}$Y) and multi-nuclide or multi-timepoint studies, where uptake patterns may differ but anatomical boundaries remain informative. In such cases, the directional/anatomical component is still well motivated, while the inter-modality coupling strength may need to be relaxed or made spatially adaptive to avoid transferring modality-specific or temporally evolving uptake features. This is an important direction for future work.

\section{Conclusion}
\label{sec:conclusion}

Post-treatment $^{90}$Y imaging for SIRT dosimetry presents a challenging reconstruction problem: PET offers high spatial resolution but suffers from extreme noise due to the low positron branching ratio, whilst SPECT provides higher count rates but exhibits poor resolution due to the continuous bremsstrahlung spectrum. To exploit the fact that both PET and SPECT share the same underlying distribution, this work introduces weighted directional total nuclear variation (\wdtnv), a synergistic regulariser for joint PET/SPECT reconstruction that couples gradient orientations between modalities via the nuclear norm of their joint Jacobian.  CT anatomical information is incorporated as a directional weighting operator.

Validation on a NEMA IEC phantom with 20 bootstrapped noise realisations demonstrated that \wdtnv achieved significantly higher recovery coefficients than both the isotropic variant (\wtnv) and non-synergistic \dtv. It also showed improved recovery for small VOIs relative to the current state-of-the-art method, sequential hybrid kernel expectation maximisation (SHKEM). In addition, voxel-wise and VOI-mean repeatability across noise realisations improved compared with SHKEM in the background and high-count regions, although repeatability was reduced relative to \wtnv and \dtv and SHKEM remained most stable in the cold lung insert. Clinical validation across 45 lesions in 9 patients revealed heterogeneous lesion-specific responses to directional guidance: lesions with well-defined anatomical correlates (e.g.\ calcification, liver boundaries) benefited substantially from synergy and \mumap-informed edge alignment, whereas lesions without corresponding anatomical features showed minimal improvement. At approximately matched background coefficient of variation (CoV), \wdtnv also achieved a significantly higher tumour-to-background ratio (TBR) than \dtv, \wtnv, and SHKEM in the full patient cohort.

This work demonstrates that synergistic reconstruction with directional \mumap guidance can improve noise-contrast trade-offs in $^{90}$Y imaging. The observed improvements in recovery/contrast--noise trade-offs and repeatability suggest potential to improve the stability of activity estimates used in post-treatment verification. However, careful consideration of CT artefacts and registration will be necessary for clinical translation.

\section*{Code Availability}
Code and configuration files used to generate the results will be made available upon publication.

\bibliographystyle{IEEEtran}
\bibliography{LaTeX/references}

@article{porter_simultaneous_2025,
	title = {Simultaneous {Multi}-{Bed} {MAP} {Reconstruction} with {CT}-{Guided} {Directional} {TV} {Prior} for {Y}-90 {PET} {SIRT}},
	volume = {2},
	copyright = {Copyright (c) 2025 Sam Porter, Daniel Deidda, Daniel McGowan, Simon Arridge, Kris Thielemans},
	issn = {3049-8902},
	url = {https://journals.ed.ac.uk/PiWJournal/article/view/10841},
	doi = {10.2218/piwjournal.10841},
	language = {en},
	number = {1},
	urldate = {2026-04-25},
	journal = {The PET is Wonderful Journal},
	author = {Porter, Sam and Deidda, Daniel and McGowan, Daniel and Arridge, Simon and Thielemans, Kris},
	month = oct,
	year = {2025},
}

@inproceedings{porter_optimising_2024,
	title = {Optimising {Subset} {Selection} in {Synergistic} {Emission} {Tomography} {Reconstruction}},
	issn = {2577-0829},
	url = {https://ieeexplore.ieee.org/abstract/document/10654962},
	doi = {10.1109/NSS/MIC/RTSD57108.2024.10654962},
	urldate = {2026-04-25},
	booktitle = {2024 {IEEE} {Nuclear} {Science} {Symposium} ({NSS}), {Medical} {Imaging} {Conference} ({MIC}) and {Room} {Temperature} {Semiconductor} {Detector} {Conference} ({RTSD})},
	author = {Porter, S. D. and Deidda, D. and Arridge, S. and Thielemans, K.},
	month = oct,
	year = {2024},
	note = {ISSN: 2577-0829},
	pages = {1--2},
}

@article{wasserthal_totalsegmentator_2023,
	title = {{TotalSegmentator}: {Robust} {Segmentation} of 104 {Anatomic} {Structures} in {CT} {Images}},
	volume = {5},
	issn = {2638-6100},
	shorttitle = {{TotalSegmentator}},
	url = {https://pmc.ncbi.nlm.nih.gov/articles/PMC10546353/},
	doi = {10.1148/ryai.230024},
	number = {5},
	urldate = {2026-04-25},
	journal = {Radiology: Artificial Intelligence},
	author = {Wasserthal, Jakob and Breit, Hanns-Christian and Meyer, Manfred T. and Pradella, Maurice and Hinck, Daniel and Sauter, Alexander W. and Heye, Tobias and Boll, Daniel T. and Cyriac, Joshy and Yang, Shan and Bach, Michael and Segeroth, Martin},
	month = jul,
	year = {2023},
	pages = {e230024},
}

@incollection{pinheiro_fitting_2000-1,
	address = {New York, NY},
	title = {Fitting {Linear} {Mixed}-{Effects} {Models}},
	isbn = {978-0-387-22747-4},
	url = {https://doi.org/10.1007/0-387-22747-4_4},
	doi = {10.1007/0-387-22747-4_4},
	language = {en},
	urldate = {2026-04-25},
	booktitle = {Mixed-{Effects} {Models} in {S} and {S}-{PLUS}},
	publisher = {Springer},
	editor = {Pinheiro, José C. and Bates, Douglas M.},
	year = {2000},
	pages = {133--199},
}

@article{laird_random-effects_1982,
	title = {Random-effects models for longitudinal data},
	volume = {38},
	issn = {0006-341X},
	language = {eng},
	number = {4},
	journal = {Biometrics},
	author = {Laird, N. M. and Ware, J. H.},
	month = dec,
	year = {1982},
	pages = {963--974},
}

@article{benjaminiControllingFalseDiscovery1995,
	title = {Controlling the {False} {Discovery} {Rate}: {A} {Practical} and {Powerful} {Approach} to {Multiple} {Testing}},
	volume = {57},
	issn = {0035-9246},
	shorttitle = {Controlling the {False} {Discovery} {Rate}},
	url = {https://www.jstor.org/stable/2346101},
	number = {1},
	urldate = {2026-04-19},
	journal = {Journal of the Royal Statistical Society. Series B (Methodological)},
	publisher = {[Royal Statistical Society, Oxford University Press]},
	author = {Benjamini, Yoav and Hochberg, Yosef},
	year = {1995},
	pages = {289--300},
}

@article{rowleyOptimizationImageReconstruction2016,
	chapter = {Clinical Investigations (Human)},
	title = {Optimization of image reconstruction for yttrium-90 {SIRT} on a {LYSO} {PET}/{CT} system using a {Bayesian} penalized likelihood reconstruction algorithm},
	copyright = {Copyright © 2016 by the Society of Nuclear Medicine and Molecular Imaging, Inc.},
	issn = {0161-5505, 2159-662X},
	url = {https://jnm.snmjournals.org/content/early/2016/09/28/jnumed.116.176552},
	doi = {10.2967/jnumed.116.176552},
	language = {en},
	urldate = {2026-02-12},
	journal = {Journal of Nuclear Medicine},
	publisher = {Society of Nuclear Medicine},
	author = {Rowley, Lisa M. and Bradley, Kevin M. and Boardman, Philip and Hallam, Aida and McGowan, Daniel R.},
	month = sep,
	year = {2016},
}

@article{schrammPARALLELPROJOpensourceFramework2024,
	title = {{PARALLELPROJ}—an open-source framework for fast calculation of projections in tomography},
	volume = {3},
	issn = {2673-8880},
	url = {https://www.frontiersin.org/journals/nuclear-medicine/articles/10.3389/fnume.2023.1324562/full},
	doi = {10.3389/fnume.2023.1324562},
	language = {English},
	urldate = {2026-01-26},
	journal = {Frontiers in Nuclear Medicine},
	publisher = {Frontiers},
	author = {Schramm, Georg and Thielemans, Kris},
	month = jan,
	year = {2024},
}

@inproceedings{johnsonAcceleratingStochasticGradient2013,
	title = {Accelerating {Stochastic} {Gradient} {Descent} using {Predictive} {Variance} {Reduction}},
	volume = {26},
	urldate = {2026-01-12},
	booktitle = {Advances in {Neural} {Information} {Processing} {Systems}},
	publisher = {Curran Associates, Inc.},
	author = {Johnson, Rie and Zhang, Tong},
	year = {2013},
}

@article{paretoNewTheoriesEconomics1897,
	title = {The {New} {Theories} of {Economics}},
	copyright = {Copyright 1897 The University of Chicago},
	issn = {0022-3808},
	url = {https://www.journals.uchicago.edu/doi/10.1086/250454},
	doi = {10.1086/250454},
	language = {en},
	urldate = {2026-01-11},
	journal = {Journal of Political Economy},
	publisher = {The University of Chicago},
	author = {Pareto, Vilfredo},
	month = sep,
	year = {1897},
}

@article{pasciakRadioembolizationDynamicRole2014,
	title = {Radioembolization and the {Dynamic} {Role} of {90Y} {PET}/{CT}},
	volume = {4},
	issn = {2234-943X},
	url = {https://www.frontiersin.org/journals/oncology/articles/10.3389/fonc.2014.00038/full},
	doi = {10.3389/fonc.2014.00038},
	language = {English},
	urldate = {2026-01-08},
	journal = {Frontiers in Oncology},
	publisher = {Frontiers},
	author = {Pasciak, Alexander S. and Bourgeois, Austin C. and McKinney, J. Mark and Chang, Ted T. and Osborne, Dustin R. and Acuff, Shelley N. and Bradley, Yong C.},
	month = feb,
	year = {2014},
}

@article{mikellImpact90YPET2018,
	title = {Impact of {90Y} {PET} gradient-based tumor segmentation on voxel-level dosimetry in liver radioembolization},
	volume = {5},
	issn = {2197-7364},
	url = {https://doi.org/10.1186/s40658-018-0230-y},
	doi = {10.1186/s40658-018-0230-y},
	language = {en},
	number = {1},
	urldate = {2026-01-03},
	journal = {EJNMMI Physics},
	author = {Mikell, Justin K. and Kaza, Ravi K. and Roberson, Peter L. and Younge, Kelly C. and Srinivasa, Ravi N. and Majdalany, Bill S. and Cuneo, Kyle C. and Owen, Dawn and Devasia, Theresa and Schipper, Matthew J. and Dewaraja, Yuni K.},
	month = nov,
	year = {2018},
	pages = {31},
}

@inproceedings{iatrou3DImplementationScatter2006,
	title = {{3D} implementation of {Scatter} {Estimation} in {3D} {PET}},
	volume = {4},
	issn = {1082-3654},
	url = {https://ieeexplore.ieee.org/document/4179452},
	doi = {10.1109/NSSMIC.2006.354338},
	urldate = {2025-12-09},
	booktitle = {2006 {IEEE} {Nuclear} {Science} {Symposium} {Conference} {Record}},
	author = {Iatrou, M. and Manjeshwar, R. M. and Ross, S. G. and Thielemans, K. and Stearns, C. W.},
	month = oct,
	year = {2006},
	pages = {2142--2145},
}

@inproceedings{stearnsRandomCoincidenceEstimation2003,
	title = {Random coincidence estimation from single event rates on the {Discovery} {ST} {PET}/{CT} scanner},
	volume = {5},
	issn = {1082-3654},
	url = {https://ieeexplore.ieee.org/document/1352545},
	doi = {10.1109/NSSMIC.2003.1352545},
	urldate = {2025-12-09},
	booktitle = {2003 {IEEE} {Nuclear} {Science} {Symposium}. {Conference} {Record}},
	author = {Stearns, C.W. and McDaniel, D.L. and Kohlmyer, S.G. and Arul, P.R. and Geiser, B.P. and Shanmugam, V.},
	month = oct,
	year = {2003},
	pages = {3067--3069 Vol.5},
}

@article{mehranianNonconvexJointsparsityRegularization2016,
	chapter = {Instrumentation \&amp; Data Analysis Track},
	title = {Non-convex joint-sparsity regularization for synergistic {PET} and {SENSE} {MRI} reconstruction},
	volume = {57},
	copyright = {© 2016},
	issn = {0161-5505, 2159-662X},
	url = {https://jnm.snmjournals.org/content/57/supplement_2/639},
	language = {en},
	number = {supplement 2},
	urldate = {2025-12-04},
	journal = {Journal of Nuclear Medicine},
	publisher = {Society of Nuclear Medicine},
	author = {Mehranian, Abolfazl and Reader, Andrew},
	month = may,
	year = {2016},
	pages = {639--639},
}

@article{sheppMaximumLikelihoodReconstruction1982,
	title = {Maximum likelihood reconstruction for emission tomography},
	volume = {1},
	issn = {0278-0062},
	doi = {10.1109/TMI.1982.4307558},
	language = {eng},
	number = {2},
	journal = {IEEE transactions on medical imaging},
	author = {Shepp, L. A. and Vardi, Y.},
	year = {1982},
	pages = {113--122},
}

@article{jaskowiakInfluenceReconstructionIterations2005,
	chapter = {Clinical Investigations},
	title = {Influence of {Reconstruction} {Iterations} on {18F}-{FDG} {PET}/{CT} {Standardized} {Uptake} {Values}},
	volume = {46},
	copyright = {THE JOURNAL OF NUCLEAR MEDICINE},
	issn = {0161-5505, 2159-662X},
	url = {https://jnm.snmjournals.org/content/46/3/424},
	language = {en},
	number = {3},
	urldate = {2025-12-04},
	journal = {Journal of Nuclear Medicine},
	publisher = {Society of Nuclear Medicine},
	author = {Jaskowiak, Chris J. and Bianco, Jesus A. and Perlman, Scott B. and Fine, Jason P.},
	month = mar,
	year = {2005},
	pages = {424--428},
}

@article{veklerovStoppingRuleMLE1987,
	title = {Stopping {Rule} for the {MLE} {Algorithm} {Based} on {Statistical} {Hypothesis} {Testing}},
	volume = {6},
	issn = {1558-254X},
	url = {https://ieeexplore.ieee.org/abstract/document/4307849},
	doi = {10.1109/TMI.1987.4307849},
	number = {4},
	urldate = {2025-12-04},
	journal = {IEEE Transactions on Medical Imaging},
	author = {Veklerov, Eugene and Llacer, Jorge},
	month = dec,
	year = {1987},
	pages = {313--319},
}

@misc{ehrhardtFastPETReconstruction2025,
	title = {Fast {PET} {Reconstruction} with {Variance} {Reduction} and {Prior}-{Aware} {Preconditioning}},
	url = {http://arxiv.org/abs/2506.04976},
	doi = {10.48550/arXiv.2506.04976},
	urldate = {2025-11-25},
	publisher = {arXiv},
	author = {Ehrhardt, Matthias J. and Kereta, Zeljko and Schramm, Georg},
	month = jun,
	year = {2025},
	note = {Number: arXiv:2506.04976
arXiv:2506.04976 [physics]},
}

@inproceedings{ahnGloballyConvergentOrdered2002,
	title = {Globally convergent ordered subsets algorithms: application to tomography},
	isbn = {978-0-7803-7324-2},
	shorttitle = {Globally convergent ordered subsets algorithms},
	url = {http://ieeexplore.ieee.org/document/1009736/},
	doi = {10.1109/NSSMIC.2001.1009736},
	language = {en},
	booktitle = {2001 {IEEE} {Nuclear} {Science} {Symposium} {Conference} {Record} ({Cat}. {No}.{01CH37310})},
	publisher = {IEEE},
	author = {Ahn, Sangtae and Fessler, J A},
	year = {2002},
	pages = {1064--1068},
}

@article{ehrhardtMultimodalityImagingStructurepromoting2020a,
	title = {Multi-modality imaging with structure-promoting regularisers},
	url = {http://arxiv.org/abs/2007.11689},
	journal = {arXiv:2007.11689 [cs, eess, math]},
	author = {Ehrhardt, Matthias J},
	year = {2020},
}

@article{markiewiczAssessmentBootstrapResampling2014,
	title = {Assessment of bootstrap resampling performance for {PET} data},
	volume = {60},
	issn = {0031-9155},
	url = {https://iopscience.iop.org/article/10.1088/0031-9155/60/1/279},
	doi = {10.1088/0031-9155/60/1/279},
	number = {1},
	urldate = {2025-09-12},
	journal = {Physics in Medicine \& Biology},
	publisher = {IOP Publishing},
	author = {Markiewicz, P. J. and Reader, A. J. and Matthews, J. C.},
	month = dec,
	year = {2014},
	pages = {279},
}

@article{hasteCorrelationTechnetium99mMacroaggregated2017,
	title = {Correlation of {Technetium}-99m {Macroaggregated} {Albumin} and {Yttrium}-90 {Glass} {Microsphere} {Biodistribution} in {Hepatocellular} {Carcinoma}: {A} {Retrospective} {Review} of {Pretreatment} {Single} {Photon} {Emission} {CT} and {Posttreatment} {Positron} {Emission} {Tomography}/{CT}},
	volume = {28},
	issn = {1051-0443},
	doi = {10.1016/J.JVIR.2016.12.1221},
	number = {5},
	urldate = {2025-08-21},
	journal = {Journal of Vascular and Interventional Radiology},
	publisher = {Elsevier},
	author = {Haste, Paul and Tann, Mark and Persohn, Scott and LaRoche, Thomas and Aaron, Vasantha and Mauxion, Thibault and Chauhan, Nikhil and Dreher, Matthew R. and Johnson, Matthew S.},
	month = may,
	year = {2017},
	pages = {722--730.e1},
}

@article{selwynNewInternalPair2007,
	title = {A new internal pair production branching ratio of {90Y}: {The} development of a non-destructive assay for {90Y} and {90Sr}},
	volume = {65},
	issn = {0969-8043},
	doi = {10.1016/J.APRADISO.2006.08.009},
	number = {3},
	urldate = {2025-08-21},
	journal = {Applied Radiation and Isotopes},
	publisher = {Pergamon},
	author = {Selwyn, R. G. and Nickles, R. J. and Thomadsen, B. R. and DeWerd, L. A. and Micka, J. A.},
	month = mar,
	year = {2007},
	pages = {318--327},
}

@article{angerSensitivityResolutionLinearity1966,
	title = {Sensitivity, resolution, and linearity of the scintillation camera},
	volume = {NS-13},
	issn = {15581578},
	doi = {10.1109/TNS.1966.4324123},
	number = {3},
	urldate = {2025-08-07},
	journal = {IEEE Transactions on Nuclear Science},
	author = {Anger, Hal O.},
	year = {1966},
	pages = {380--392},
}

@article{lewisDerivativesSpectralFunctions1996,
	title = {Derivatives of {Spectral} {Functions}},
	volume = {21},
	issn = {0364-765X},
	url = {https://pubsonline.informs.org/doi/10.1287/moor.21.3.576},
	doi = {10.1287/moor.21.3.576},
	number = {3},
	urldate = {2025-06-05},
	journal = {Mathematics of Operations Research},
	publisher = {INFORMS},
	author = {Lewis, A. S.},
	month = aug,
	year = {1996},
	pages = {576--588},
}

@article{raultFastSimulationYttrium902010,
	title = {Fast simulation of yttrium-90 bremsstrahlung photons with {GATE}},
	volume = {37},
	issn = {2473-4209},
	url = {https://onlinelibrary.wiley.com/doi/abs/10.1118/1.3431998},
	doi = {10.1118/1.3431998},
	language = {en},
	number = {6Part1},
	urldate = {2024-12-11},
	journal = {Medical Physics},
	author = {Rault, Erwann and Staelens, Steven and Van Holen, Roel and De Beenhouwer, Jan and Vandenberghe, Stefaan},
	year = {2010},
	note = {\_eprint: https://onlinelibrary.wiley.com/doi/pdf/10.1118/1.3431998},
	pages = {2943--2950},
}

@article{levillainInternationalRecommendationsPersonalised2021,
	title = {International recommendations for personalised selective internal radiation therapy of primary and metastatic liver diseases with yttrium-90 resin microspheres},
	volume = {48},
	issn = {1619-7089},
	url = {https://doi.org/10.1007/s00259-020-05163-5},
	doi = {10.1007/s00259-020-05163-5},
	language = {en},
	number = {5},
	urldate = {2024-12-09},
	journal = {European Journal of Nuclear Medicine and Molecular Imaging},
	author = {Levillain, Hugo and Bagni, Oreste and Deroose, Christophe M. and Dieudonné, Arnaud and Gnesin, Silvano and Grosser, Oliver S. and Kappadath, S. Cheenu and Kennedy, Andrew and Kokabi, Nima and Liu, David M. and Madoff, David C. and Mahvash, Armeen and Martinez de la Cuesta, Antonio and Ng, David C. E. and Paprottka, Philipp M. and Pettinato, Cinzia and Rodríguez-Fraile, Macarena and Salem, Riad and Sangro, Bruno and Strigari, Lidia and Sze, Daniel Y. and de Wit van der veen, Berlinda J. and Flamen, Patrick},
	month = may,
	year = {2021},
	pages = {1570--1584},
}

@article{depierroFastEMlikeMethods2001,
	title = {Fast {EM}-like methods for maximum "a posteriori" estimates in emission tomography},
	volume = {20},
	issn = {02780062},
	url = {http://ieeexplore.ieee.org/document/921477/},
	doi = {10.1109/42.921477},
	language = {en},
	number = {4},
	urldate = {2024-03-13},
	journal = {IEEE Transactions on Medical Imaging},
	author = {De Pierro, A.R. and Yamagishi, M.E.B.},
	month = apr,
	year = {2001},
	pages = {280--288},
}

@article{deiddaTripleModalityImage2023,
	title = {Triple modality image reconstruction of {PET} data using {SPECT}, {PET}, {CT} information increases lesion uptake in images of patients treated with radioembolization with {90Y} micro-spheres},
	volume = {10},
	issn = {2197-7364},
	url = {https://doi.org/10.1186/s40658-023-00549-4},
	doi = {10.1186/s40658-023-00549-4},
	number = {1},
	urldate = {2023-05-26},
	journal = {EJNMMI Physics},
	author = {Deidda, Daniel and Denis-Bacelar, Ana M. and Fenwick, Andrew J. and Ferreira, Kelley M. and Heetun, Warda and Hutton, Brian F. and McGowan, Daniel R. and Robinson, Andrew P. and Scuffham, James and Thielemans, Kris and Twyman, Robert},
	month = may,
	year = {2023},
	pages = {30},
}

@article{rongDevelopmentEvaluationImproved2012,
	title = {Development and evaluation of an improved quantitative (90){Y} bremsstrahlung {SPECT} method},
	volume = {39},
	issn = {0094-2405},
	doi = {10.1118/1.3700174},
	language = {eng},
	number = {5},
	journal = {Medical Physics},
	author = {Rong, Xing and Du, Yong and Ljungberg, Michael and Rault, Erwann and Vandenberghe, Stefaan and Frey, Eric C.},
	month = may,
	year = {2012},
	pages = {2346--2358},
}

@article{ljungbergSIMINDMonteCarlo2015,
	chapter = {Instrumentation \&amp; Data Analysis},
	title = {{SIMIND} {Monte} {Carlo} based image reconstruction},
	volume = {56},
	issn = {0161-5505, 2159-662X},
	url = {https://jnm.snmjournals.org/content/56/supplement_3/43},
	language = {en},
	number = {supplement 3},
	urldate = {2023-01-30},
	journal = {Journal of Nuclear Medicine},
	publisher = {Society of Nuclear Medicine},
	author = {Ljungberg, Michael},
	month = may,
	year = {2015},
	pages = {43--43},
}

@article{fusterIntegrationAdvanced3D2013,
	title = {Integration of advanced {3D} {SPECT} modeling into the open-source {STIR} framework},
	volume = {40},
	issn = {2473-4209},
	doi = {10.1118/1.4816676},
	language = {eng},
	number = {9},
	journal = {Medical Physics},
	author = {Fuster, Berta Marti and Falcon, Carles and Tsoumpas, Charalampos and Livieratos, Lefteris and Aguiar, Pablo and Cot, Albert and Ros, Domenec and Thielemans, Kris},
	month = sep,
	year = {2013},
	pages = {092502},
}

@article{thielemansSTIRSoftwareTomographic2012,
	title = {{STIR}: {Software} for tomographic image reconstruction release 2},
	volume = {57},
	shorttitle = {{STIR}},
	doi = {10.1088/0031-9155/57/4/867},
	journal = {Physics in medicine and biology},
	author = {Thielemans, Kris and Tsoumpas, Charalampos and Mustafovic, Sanida and Beisel, Tobias and Aguiar, Pablo and Dikaios, Nikolaos and Jacobson, Matthew},
	month = feb,
	year = {2012},
	pages = {867--83},
}

@article{xiangDeepNeuralNetwork2020,
	title = {A deep neural network for fast and accurate scatter estimation in quantitative {SPECT}/{CT} under challenging scatter conditions},
	volume = {47},
	issn = {1619-7089},
	url = {https://doi.org/10.1007/s00259-020-04840-9},
	doi = {10.1007/s00259-020-04840-9},
	language = {en},
	number = {13},
	urldate = {2023-01-27},
	journal = {European Journal of Nuclear Medicine and Molecular Imaging},
	author = {Xiang, Haowei and Lim, Hongki and Fessler, Jeffrey A. and Dewaraja, Yuni K.},
	month = dec,
	year = {2020},
	pages = {2956--2967},
}

@article{minarikEvaluationQuantitative90Y2008,
	title = {Evaluation of quantitative {90Y} {SPECT} based on experimental phantom studies},
	volume = {53},
	issn = {0031-9155},
	url = {https://dx.doi.org/10.1088/0031-9155/53/20/008},
	doi = {10.1088/0031-9155/53/20/008},
	language = {en},
	number = {20},
	urldate = {2023-01-27},
	journal = {Physics in Medicine \& Biology},
	author = {Minarik, D. and Gleisner, K. Sjögreen and Ljungberg, M.},
	month = sep,
	year = {2008},
	pages = {5689},
}

@article{deiddaHybridKernelisedExpectation2022,
	title = {Hybrid kernelised expectation maximisation for {Bremsstrahlung} {SPECT} reconstruction in {SIRT} with {90Y} micro-spheres},
	volume = {9},
	issn = {2197-7364},
	url = {https://doi.org/10.1186/s40658-022-00452-4},
	doi = {10.1186/s40658-022-00452-4},
	language = {en},
	number = {1},
	urldate = {2023-01-27},
	journal = {EJNMMI Physics},
	author = {Deidda, Daniel and Denis-Bacelar, Ana M. and Fenwick, Andrew J. and Ferreira, Kelley M. and Heetun, Warda and Hutton, Brian F. and Robinson, Andrew P. and Scuffham, James and Thielemans, Kris},
	month = apr,
	year = {2022},
	pages = {25},
}

@article{rigieJointReconstructionMultichannel2015,
	title = {Joint {Reconstruction} of {Multi}-channel, {Spectral} {CT} {Data} via {Constrained} {Total} {Nuclear} {Variation} {Minimization}},
	volume = {60},
	issn = {0031-9155},
	url = {https://www.ncbi.nlm.nih.gov/pmc/articles/PMC4669200/},
	doi = {10.1088/0031-9155/60/5/1741},
	number = {5},
	urldate = {2022-07-06},
	journal = {Physics in medicine and biology},
	author = {Rigie, David S. and La Rivière, Patrick J.},
	month = mar,
	year = {2015},
	pages = {1741--1762},
}

@inproceedings{ovtchinnikovSIRFSynergisticImage2017,
	address = {Atlanta, GA},
	title = {{SIRF}: {Synergistic} {Image} {Reconstruction} {Framework}},
	isbn = {978-1-5386-2282-7},
	shorttitle = {{SIRF}},
	url = {https://ieeexplore.ieee.org/document/8532815/},
	doi = {10.1109/NSSMIC.2017.8532815},
	urldate = {2022-05-03},
	booktitle = {2017 {IEEE} {Nuclear} {Science} {Symposium} and {Medical} {Imaging} {Conference} ({NSS}/{MIC})},
	publisher = {IEEE},
	author = {Ovtchinnikov, Evgueni and Atkinson, David and Kolbitsch, Christoph and Thomas, Benjamin A. and Bertolli, Ottavia and da Costa-Luis, Casper O. and Efthimiou, Nikolaos and Fowler, Ronald and Pasca, Edoardo and Wadhwa, Palak and Emond, Elise and Matthews, Julian and Prieto, Claudia and Reader, Andrew J. and Tsoumpas, Charalampos and Turner, Martin and Thielemans, Kris},
	month = oct,
	year = {2017},
	pages = {1--3},
}

@article{jorgensenCoreImagingLibrary2021,
	title = {Core {Imaging} {Library} - {Part} {I}: a versatile {Python} framework for tomographic imaging},
	volume = {379},
	issn = {1364-503X, 1471-2962},
	shorttitle = {Core {Imaging} {Library} - {Part} {I}},
	url = {https://royalsocietypublishing.org/doi/10.1098/rsta.2020.0192},
	doi = {10.1098/rsta.2020.0192},
	language = {en},
	number = {2204},
	urldate = {2022-03-08},
	journal = {Philosophical Transactions of the Royal Society A: Mathematical, Physical and Engineering Sciences},
	author = {Jørgensen, J. S. and Ametova, E. and Burca, G. and Fardell, G. and Papoutsellis, E. and Pasca, E. and Thielemans, K. and Turner, M. and Warr, R. and Lionheart, W. R. B. and Withers, P. J.},
	month = aug,
	year = {2021},
	pages = {20200192},
}

@article{knollJointMRPETReconstruction2017,
	title = {Joint {MR}-{PET} {Reconstruction} {Using} a {Multi}-{Channel} {Image} {Regularizer}},
	volume = {36},
	issn = {1558-254X},
	doi = {10.1109/TMI.2016.2564989},
	number = {1},
	journal = {IEEE Transactions on Medical Imaging},
	author = {Knoll, Florian and Holler, Martin and Koesters, Thomas and Otazo, Ricardo and Bredies, Kristian and Sodickson, Daniel K},
	month = jan,
	year = {2017},
	note = {Conference Name: IEEE Transactions on Medical Imaging},
	pages = {1--16},
}

@inproceedings{bowsherUtilizingMRIInformation2004,
	title = {Utilizing {MRI} information to estimate {F18}-{FDG} distributions in rat flank tumors},
	volume = {4},
	issn = {1082-3654},
	doi = {10.1109/NSSMIC.2004.1462760},
	booktitle = {{IEEE} {Symposium} {Conference} {Record} {Nuclear} {Science} 2004.},
	author = {Bowsher, J.E. and Yuan, Hong and Hedlund, L.W. and Turkington, T.G. and Akabani, G. and Badea, A. and Kurylo, W.C. and Wheeler, C.T. and Cofer, G.P. and Dewhirst, M.W. and Johnson, G.A.},
	month = oct,
	year = {2004},
	pages = {2488--2492 Vol. 4},
}

@article{marquisTheranosticSPECTReconstruction2021,
	title = {Theranostic {SPECT} reconstruction for improved resolution: application to radionuclide therapy dosimetry},
	volume = {8},
	issn = {2197-7364},
	shorttitle = {Theranostic {SPECT} reconstruction for improved resolution},
	url = {https://ejnmmiphys.springeropen.com/articles/10.1186/s40658-021-00362-x},
	doi = {10.1186/s40658-021-00362-x},
	language = {en},
	number = {1},
	urldate = {2022-02-09},
	journal = {EJNMMI Physics},
	author = {Marquis, H. and Deidda, D. and Gillman, A. and Willowson, K. P. and Gholami, Y. and Hioki, T. and Eslick, E. and Thielemans, K. and Bailey, D. L.},
	month = dec,
	year = {2021},
	pages = {16},
}

@article{dempsterMaximumLikelihoodIncomplete1977,
	title = {Maximum {Likelihood} from {Incomplete} {Data} {Via} the {EM} {Algorithm}},
	volume = {39},
	issn = {00359246},
	url = {https://onlinelibrary.wiley.com/doi/10.1111/j.2517-6161.1977.tb01600.x},
	doi = {10.1111/j.2517-6161.1977.tb01600.x},
	language = {en},
	number = {1},
	urldate = {2021-11-18},
	journal = {Journal of the Royal Statistical Society: Series B (Methodological)},
	author = {Dempster, A. P. and Laird, N. M. and Rubin, D. B.},
	month = sep,
	year = {1977},
	pages = {1--22},
}

@article{arridgeOverviewSynergisticReconstruction2021,
	title = {({An} overview of) {Synergistic} reconstruction for multimodality/multichannel imaging methods},
	volume = {379},
	url = {https://royalsocietypublishing.org/doi/abs/10.1098/rsta.2020.0205},
	doi = {10.1098/RSTA.2020.0205},
	number = {2200},
	journal = {Philosophical Transactions of the Royal Society A},
	publisher = {The Royal Society Publishing},
	author = {Arridge, Simon R. and Ehrhardt, Matthias J. and Thielemans, Kris},
	month = jun,
	year = {2021},
}

@article{hudsonAcceleratedImageReconstruction1994,
	title = {Accelerated image reconstruction using ordered subsets of projection data},
	volume = {13},
	issn = {1558-254X},
	doi = {10.1109/42.363108},
	number = {4},
	journal = {IEEE Transactions on Medical Imaging},
	author = {Hudson, H.M. and Larkin, R.S.},
	month = dec,
	year = {1994},
	note = {Conference Name: IEEE Transactions on Medical Imaging},
	pages = {601--609},
}

@article{deiddaHybridPETMRListmode2019,
	title = {Hybrid {PET}-{MR} list-mode kernelized expectation maximization reconstruction},
	volume = {35},
	issn = {0266-5611, 1361-6420},
	url = {https://iopscience.iop.org/article/10.1088/1361-6420/ab013f},
	doi = {10.1088/1361-6420/ab013f},
	language = {en},
	number = {4},
	urldate = {2021-11-26},
	journal = {Inverse Problems},
	author = {Deidda, Daniel and Karakatsanis, Nicolas A and Robson, Philip M and Tsai, Yu-Jung and Efthimiou, Nikos and Thielemans, Kris and Fayad, Zahi A and Aykroyd, Robert G and Tsoumpas, Charalampos},
	month = apr,
	year = {2019},
	pages = {044001},
}

@article{hutchcroftAnatomicallyaidedPETReconstruction2016,
	title = {Anatomically-aided {PET} reconstruction using the kernel method},
	volume = {61},
	issn = {0031-9155, 1361-6560},
	url = {https://iopscience.iop.org/article/10.1088/0031-9155/61/18/6668},
	doi = {10.1088/0031-9155/61/18/6668},
	language = {en},
	number = {18},
	urldate = {2021-11-26},
	journal = {Physics in Medicine and Biology},
	author = {Hutchcroft, Will and Wang, Guobao and Chen, Kevin T and Catana, Ciprian and Qi, Jinyi},
	month = sep,
	year = {2016},
	pages = {6668--6683},
}

@article{wangPETImageReconstruction2015,
	title = {{PET} {Image} {Reconstruction} {Using} {Kernel} {Method}},
	volume = {34},
	issn = {1558-254X},
	doi = {10.1109/TMI.2014.2343916},
	number = {1},
	journal = {IEEE Transactions on Medical Imaging},
	author = {Wang, Guobao and Qi, Jinyi},
	month = jan,
	year = {2015},
	note = {Conference Name: IEEE Transactions on Medical Imaging},
	pages = {61--71},
}

@misc{jonssonTotalVariationRegularizationPositron1998,
	title = {Total-{Variation} {Regularization} in {Positron} {Emission} {Tomography}},
	author = {Jonsson, Elias and Huang, Sung-cheng and Chan, Tony},
	year = {1998},
}

@article{ehrhardtPETReconstructionAnatomical2016,
	title = {{PET} {Reconstruction} {With} an {Anatomical} {MRI} {Prior} {Using} {Parallel} {Level} {Sets}},
	volume = {35},
	issn = {1558-254X},
	doi = {10.1109/TMI.2016.2549601},
	number = {9},
	journal = {IEEE Transactions on Medical Imaging},
	author = {Ehrhardt, Matthias J. and Markiewicz, Pawel and Liljeroth, Maria and Barnes, Anna and Kolehmainen, Ville and Duncan, John S. and Pizarro, Luis and Atkinson, David and Hutton, Brian F. and Ourselin, Sébastien and Thielemans, Kris and Arridge, Simon R.},
	month = sep,
	year = {2016},
	note = {Conference Name: IEEE Transactions on Medical Imaging},
	pages = {2189--2199},
}

@article{holtTotalNuclearVariation2014,
	title = {Total {Nuclear} {Variation} and {Jacobian} {Extensions} of {Total} {Variation} for {Vector} {Fields}},
	volume = {23},
	issn = {1941-0042},
	doi = {10.1109/TIP.2014.2332397},
	number = {9},
	journal = {IEEE Transactions on Image Processing},
	author = {Holt, Kevin M.},
	month = sep,
	year = {2014},
	note = {Conference Name: IEEE Transactions on Image Processing},
	pages = {3975--3989},
}

@article{rudinNonlinearTotalVariation1992,
	title = {Nonlinear total variation based noise removal algorithms},
	volume = {60},
	issn = {0167-2789},
	url = {https://www.sciencedirect.com/science/article/pii/016727899290242F},
	doi = {10.1016/0167-2789(92)90242-F},
	language = {en},
	number = {1},
	urldate = {2022-01-08},
	journal = {Physica D: Nonlinear Phenomena},
	author = {Rudin, Leonid I. and Osher, Stanley and Fatemi, Emad},
	month = nov,
	year = {1992},
	pages = {259--268},
}

@article{ehrhardtJointReconstructionPETMRI2015,
	title = {Joint reconstruction of {PET}-{MRI} by exploiting structural similarity},
	volume = {31},
	doi = {10.1088/0266-5611/31/1/015001},
	journal = {Inverse Problems},
	author = {Ehrhardt, Matthias and Thielemans, Kris and Pizarro, Luis and Atkinson, David and Ourselin, Sébastien and Hutton, Brian and Arridge, Simon},
	month = jan,
	year = {2015},
	pages = {015001},
}

\end{document}